\documentclass[aps,prx,reprint,superscriptaddress,showkeys,longbibliography,floatfix]{revtex4-2}

\usepackage{graphicx}
\usepackage{amssymb}
\usepackage{amsmath}
\usepackage{amsfonts}
\usepackage{booktabs}
\usepackage{siunitx}
\usepackage{mathrsfs}
\usepackage{mathtools}
\usepackage{multirow}
\usepackage{physics}
\usepackage{placeins}

\newcommand{\foreign}[1]{\textit{#1}}
\newcommand{\spqe}[1]{SPQE}
\newcommand{\adaptsd}[1]{ADAPT-VQE-SD}
\newcommand{\adaptgsd}[1]{ADAPT-VQE-GSD}

\begin{document}

\title{CNOT-Efficient Circuits for Arbitrary Rank Many-Body Fermionic and Qubit Excitations}

\author{Ilias Magoulas}
\email{ilias.magoulas@emory.edu}
\affiliation
{Department of Chemistry and Cherry Emerson Center for Scientific Computation,
	Emory University, Atlanta, Georgia 30322, USA}

\author{Francesco A.\ Evangelista}
\email{francesco.evangelista@emory.edu}
\affiliation
{Department of Chemistry and Cherry Emerson Center for Scientific Computation,
	Emory University, Atlanta, Georgia 30322, USA}

\date{\today}

\begin{abstract}

Efficient quantum circuits are necessary for realizing quantum algorithms on noisy 
intermediate-scale quantum devices. Fermionic excitations entering unitary coupled-cluster 
(UCC) ans\"{a}tze give rise to quantum circuits containing CNOT ``staircases'' whose number scales 
exponentially with the excitation rank. Recently, Yordanov \foreign{et al.}\ [\textit{Phys.\ Rev.\ 
A} \textbf{102}, 062612 (2020); \textit{Commun.\ Phys.} \textbf{4}, 228 (2021)]
constructed CNOT-efficient quantum circuits for both fermionic- (FEB) and qubit-excitation-based 
(QEB) singles and doubles and illustrated their usefulness in adaptive derivative-assembled 
pseudo-Trotterized variational quantum eigensolver (ADAPT-VQE) simulations.
In this work, we extend these CNOT-efficient 
quantum circuits to arbitrary excitation ranks. To illustrate the benefits of these compact FEB 
and 
QEB quantum circuits, we perform numerical simulations using the recently developed selected 
projective quantum eigensolver (SPQE) approach, which relies on an adaptive UCC ansatz built from 
arbitrary-order particle--hole excitation operators. We show that both FEB- and QEB-SPQE decrease 
the number of CNOT gates compared to traditional SPQE by factors as large as 15. At the same time, 
QEB-SPQE requires, in general, more ansatz parameters than FEB-SPQE, in particular those 
corresponding to higher-than-double excitations, resulting in quantum circuits with larger CNOT 
counts. Although ADAPT-VQE generates quantum circuits with fewer CNOTs than SPQE, SPQE requires 
orders of magnitude less residual element evaluations than gradient element evaluations in 
ADAPT-VQE.

\end{abstract}

\maketitle

\section{Introduction}
\label{sec_intro}

Since their inception more than 40 years ago, quantum computers have been anticipated to perform 
certain computational tasks more efficiently than a classical machine \cite{Benioff1980,Manin1980, 
Feynman1982,Preskill2021}. An area where quantum computers can potentially have an 
advantage over their classical counterparts is the simulation of quantum many-body systems, such as 
those encountered in chemistry, condensed matter physics, and materials science. One of the main 
reasons behind this speed-up is the ability of quantum devices to represent highly entangled 
many-body states with a number of qubits proportional to the system size, a manifestation 
of Feynman's rule of simulation \cite{Feynman1982}.

Unfortunately, the reality is more complicated and physical realizations of quantum devices are 
plagued by various kinds of errors, such as imperfections of qubits, limited 
connectivity among qubits, decoherence of quantum states, readout errors, and errors in 
implementing quantum gates \cite{Preskill2018}. These issues could be remedied by 
fault-tolerant quantum computers, utilizing many physical qubits to encode a single logical one 
\cite{Roffe2019}. Nevertheless, scaling up the size of quantum devices is nothing short of an 
``experimenter's nightmare" \cite{Haroche1996}.

The technological advances aimed at improving the error rates of present noisy 
intermediate-scale quantum (NISQ) devices \cite{Preskill2018} have been accompanied by developments 
in the field of quantum algorithms \cite{Bharti2022}.
To fully utilize current quantum hardware,  
hybrid quantum--classical approaches have been developed that require
shallower circuits than pure quantum algorithms, \foreign{e.g.}, quantum phase estimation 
\cite{Kitaev1995,Abrams1997,Abrams1999}. The first such hybrid approach was the variational quantum 
eigensolver (VQE) \cite{Peruzzo2014,McClean2016,Cerezo2021,Tilly2021,Fedorov2022b}, where the 
lowest eigenvalue of a Hamiltonian is estimated by minimizing its expectation value with 
respect to a parameterized trial state.
In the projective 
quantum eigensolver (PQE) \cite{Stair2021}, the 
optimization procedure relies on residuals, namely, 
projections of the Schr\"{o}dinger equation onto a linearly independent basis.  The contracted 
quantum eigensolver (CQE) \cite{Smart2021} is another approach using residuals, this time of the 
two-particle anti-Hermitian contracted Schr\"{o}dinger equation \cite{Mazziotti2006}.
In applications to electronic structure, one typically employs 
fermionic ans\"{a}tze. In 
the case of VQE and PQE, in particular, the chemically inspired ans\"{a}tze 
based on the unitary extension 
\cite{Kutzelnigg1977,Kutzelnigg1982,Kutzelnigg1983,Kutzelnigg1984,Bartlett1989,Szalay1995,Taube2006,
	Cooper2010,Evangelista2011,Harsha2018,Filip2020,Freericks2022,Anand2022}
of coupled-cluster (CC) theory \cite{Coester:1958,Coester:1960,cizek1,cizek2,cizek3,cizek4} 
(UCC) have been most popular.
Recently, ans\"{a}tze based on qubit excitations have been explored in the 
context of VQE \cite{Ryabinkin2018,Ryabinkin2020,Yordanov2020,Yordanov2021,Tang2021,Xia2021} 
and CQE \cite{Mazziotti2021,Smart2022}. Such algorithms are potentially more computationally frugal 
since qubit excitations are native to quantum computers. 

To further reduce the computational resources required by hybrid 
algorithms, extensions have been proposed that rely on 
iteratively constructed rather than fixed ans\"{a}tze \cite{Grimsley2019,Tang2021,Yordanov2021,
	Ryabinkin2020,Stair2021,Smart2021,Mazziotti2021,Smart2022,Fedorov2022}.
This additional flexibility results in a 
substantial reduction in the number of parameters needed to achieve a given 
accuracy in the computed energies, giving rise to shallower quantum circuits. In the popular 
adaptive derivative-assembled pseudo-Trotterized (ADAPT) VQE method \cite{Grimsley2019}, for 
example, one typically builds the ansatz one operator at a time from a pool containing single and 
double particle--hole excitation operators or their generalized variants 
\cite{Nooijen2000,Nakatsuji2000}. To allow ADAPT-VQE to converge to the exact, full configuration 
interaction (FCI), solution, a 
given excitation operator maybe be added to the ansatz multiple times, albeit with a different 
excitation amplitude.
In contrast to ADAPT-VQE, the selected PQE (SPQE) approach 
\cite{Stair2021} expands the ansatz by selecting a batch of operators from a complete pool of 
particle--hole excitations, including up to $N$-tuples where $N$ is the number of correlated 
electrons. In SPQE, the operator pool is allowed to ``drain'', \foreign{i.e.}, once an operator is 
selected to be appended to the ansatz, it is removed from the pool.

Despite the success of adaptive approaches, they generally still require a substantial number of 
two-qubit CNOT gates that is typically greater than that of the single-qubit ones. The fact that 
physical implementations of two-qubit gates are characterized by errors that are 
typically one order of magnitude larger than those associated with their single-qubit analogs (see, 
\foreign{e.g.}, Ref.\ \cite{google_weber})
implies that CNOTs dominate the overall gate error. Consequently, to facilitate the experimental 
realization of quantum algorithms, it is crucial to reduce the CNOT 
count without introducing drastic approximations that adversely affect accuracies.

Recently, Yordanov \foreign{et al.}\ constructed the most CNOT-efficient quantum circuits to date 
representing single and double qubit excitations.
Their qubit-excitation-based (QEB) approach reduced the CNOT 
count by 50\% in the case of qubit singles and 73\% for qubit doubles. These 
impressive 
reductions were attained without sacrificing accuracy, since the QEB 
quantum circuits are equivalent to their conventional counterparts. Yordanov \foreign{et al.}\ also 
extended their approach to fermionic excitations by making suitable modifications to their QEB 
quantum circuits. The performance of the fermionic-excitation-based (FEB) and QEB variants of 
ADAPT-VQE was recently examined \cite{Yordanov2021}. It was demonstrated that QEB-ADAPT-VQE is 
almost as accurate as its FEB counterpart while requiring fewer CNOTs. In addition, QEB-ADAPT-VQE 
was shown to yield more compact ans\"{a}tze than the qubit-ADAPT-VQE scheme of Ref.\ 
\cite{Tang2021} when higher accuracy is desired.

Encouraged by the substantial reduction in CNOT count for single and double excitations, in 
this work we extend the CNOT-efficient quantum circuits of Yordanov \foreign{et 
al.}\ \cite{Yordanov2020} to the $n$-body case. To illustrate their benefits, we implement and 
benchmark the FEB and QEB versions of the SPQE approach.
In particular, we perform classical numerical 
simulations for the symmetric dissociations of $\text{BeH}_2$ (linear) and
$\text{H}_6$ (linear, ring) and for the 
insertion of the Be atom into $\text{H}_2$. To demonstrate the savings in the CNOT count, we 
begin by comparing the FEB- and QEB-SPQE schemes with their traditional fermionic 
and qubit counterparts. We then proceed to the determination of the SPQE flavor that offers the 
best balance between accuracy and computational cost. 
Finally, we compare the best SPQE and ADAPT-VQE variants.

\section{Theory}
\label{sec_theor_comput}

\subsection{Background}
\label{sec_theory}

The ground electronic state ($\Psi_0$) and corresponding energy ($E_0$) of a many-electron 
system can be obtained by solving the electronic Schr\"{o}dinger equation,
\begin{equation}
	\label{eq_SE}
	H \ket*{\Psi_0} = E_0 \ket*{\Psi_0}.
\end{equation}
In the language of second quantization, the electronic Hamiltonian $H$ is expressed as
\begin{equation}
	\label{eq_H}
	H = \sum_{pq} h_{pq} a^p a_q + \frac{1}{4} \sum_{pqrs} v_{pqrs} a^p a^q a_s a_r,
\end{equation}
where $a^p \equiv a_p^\dagger$ ($a_p$) is the fermionic creation (annihilation) operator associated 
with spinorbital $\psi_p$ and $h_{pq}$ and $v_{pqrs}$ are one- and two-electron integrals. In 
general, 
hybrid quantum--classical approaches rely on a unitary parameterization of 
the wavefunction that can be readily implemented on a quantum device, \foreign{i.e.},
\begin{equation}
	\label{eq_trial_state}
	\ket*{\tilde{\Psi}(\mathbf{t})} = U(\mathbf{t}) \ket*{\Phi},
\end{equation}
where $\mathbf{t}$ represents a set of parameters and $\ket*{\Phi}$ is an easily prepared reference 
state. The energy of this trial state is computed as an expectation value and is guaranteed to be 
an upper bound to the exact ground-state energy $E_0$,
\begin{equation}
	\label{eq_ev}
	\ev*{U^\dagger (\mathbf{t}) H U(\mathbf{t})}{\Phi} \ge E_0,
\end{equation}
due to the variational principle. There are various ways to optimize the trial state. In VQE, for 
example, the optimum parameters are obtained by minimizing the VQE energy, namely,
\begin{equation}
	\label{eq_vqe_nrg}
	E_\text{VQE} = \min_\mathbf{t} \ev*{U^\dagger (\mathbf{t}) H U(\mathbf{t})}{\Phi}.
\end{equation}
An alternative strategy is offered by PQE, where the parameters are chosen such that a set of 
residual conditions,
\begin{equation}
	\label{eq_pqe_res_con}
	r_\mu (\mathbf{t}) \equiv \mel*{\Phi_\mu}{U^\dagger(\mathbf{t}) H U(\mathbf{t})}{\Phi} = 0,
\end{equation}
is satisfied, with $\{\ket*{\Phi_\mu}\}$ being an orthonormal set of many-electron states 
orthogonal to $\ket*{\Phi}$ (see Appendix A for the algorithmic details). Depending on whether the 
number of 
parameters remains constant throughout the optimization process or is iteratively increased, one 
can have fixed-ans\"{a}tze schemes or adaptive approaches such as ADAPT-VQE and SPQE. When 
employing a full set of parameters, both VQE and PQE solutions become equivalent to FCI.

\begin{figure*}[t]
	\centering
	\includegraphics[width=\textwidth]{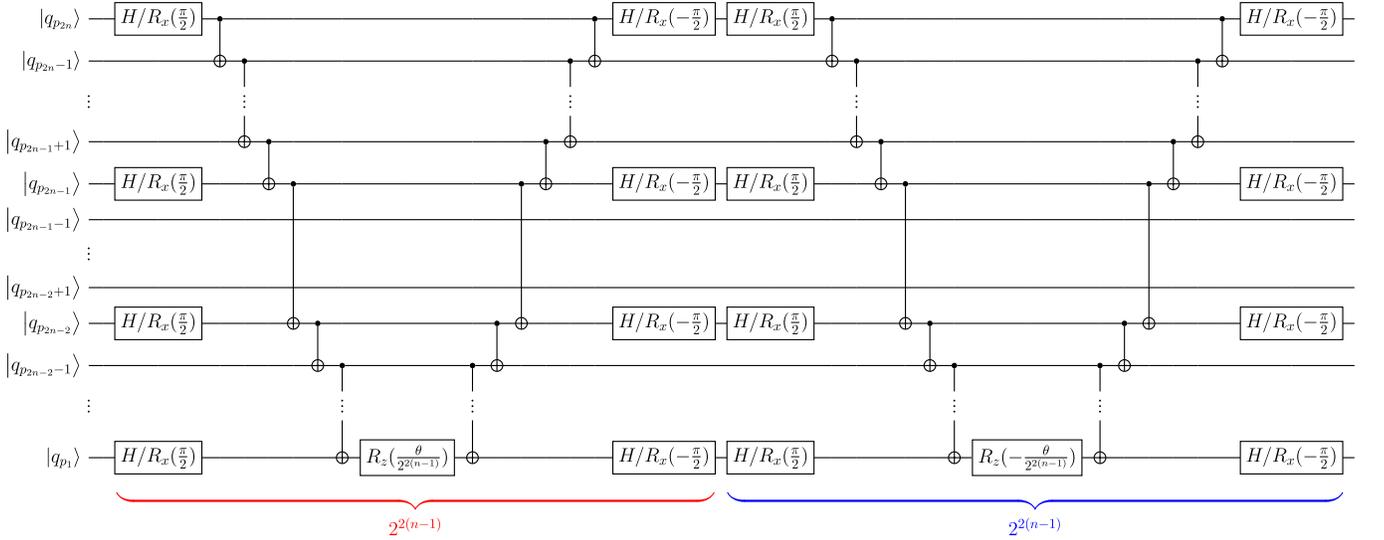}
	\caption{\label{figure1}
		Standard quantum circuit performing a fermionic $n$-tuple particle--hole excitation,
		$\exp(\theta \kappa_{p_1 \ldots p_n}^{p_{n+1} \ldots p_{2n}})$. The vertical dotted lines 
		denote CNOT ``staircases.''
	}
\end{figure*}

In the majority of quantum computing applications to chemistry, the reference state $\ket*{\Phi}$ 
appearing in Eq.\ \eqref{eq_trial_state} is a single Slater determinant, typically the 
Hartree--Fock 
(HF) determinant, 
and $U(\mathbf{t})$ is a factorized UCC unitary,
\begin{equation}
	\label{eq_ducc_unitary}
	U^{(\text{dUCC})}(\mathbf{t}) = \prod_\mu e^{t_\mu \kappa_\mu},
\end{equation}
which we refer to as disentangled UCC (dUCC) \cite{Evangelista2019}. The $\kappa_\mu$ symbols 
denote anti-Hermitian many-body excitation operators, defined as $\kappa_\mu \equiv \kappa_{p_1 
\ldots p_n}^{p_{n+1} \ldots p_{2n}} = a^{p_{n+1}} \cdots a^{p_{2n}} a_{p_n} \cdots a_{p_1} - 
a^{p_1} \cdots a^{p_n} a_{p_{2n}} \cdots a_{p_{n+1}}$. Here, the first $n$ indices, namely, $p_1 < 
\cdots < p_n$, designate spinorbitals occupied in the reference Slater determinant $\ket*{\Phi}$ 
while the $p_{n+1} < \cdots < p_{2n}$ indices denote unoccupied ones. Note that the presence of 
de-excitation operators in the UCC wavefunction ansatz implies that the $\kappa_\mu$ operators do 
not necessarily commute among themselves and, consequently, the ordering of the exponentials in 
$U^{(\text{dUCC})}(\mathbf{t})$ is important.

\subsection{Standard Fermionic and Qubit Quantum Circuits}
\label{sec_standard_circuits}

The first step in realizing a quantum chemistry calculation on a quantum computer is to employ a 
fermionic encoding, enabling one to represent a fermionic problem in the qubit basis. In the 
Jordan--Wigner (JW) transformation \cite{Jordan1928}, which we employ throughout this study, the 
fermionic creation ($a^p$)
and annihilation ($a_p$) operators associated with spinorbital $p$ are transformed as
\begin{equation}\label{eq_jw_create}
	a^p \xrightarrow{\text{JW}} Q^p \prod_{i = 0}^{p - 1} Z_i
\end{equation}
and
\begin{equation}\label{eq_jw_annihilate}
	a_p \xrightarrow{\text{JW}} Q_p \prod_{i = 0}^{p - 1} Z_i,
\end{equation}
where $Q^p \equiv Q_p^\dagger = \tfrac{1}{2} (X_p - i Y_p)$ and $Q_p = \tfrac{1}{2} (X_p + i Y_p)$ 
are the corresponding qubit creation and annihilation operators, respectively, and $X_p$, $Y_p$, 
and $Z_p$ denote Pauli gates acting on the $p^\text{th}$ qubit. Under the JW mapping, 
spinorbital occupation numbers are stored locally in the quantum device while the fermionic sign 
is encoded non-locally.

Using Eqs.\ \eqref{eq_jw_create} and \eqref{eq_jw_annihilate}, one can ``translate'' any fermionic 
operator from the language of second quantization to linear combinations of Pauli strings acting on 
qubits. For the purposes of this study, we focus on an element of the dUCC ansatz, Eq.\ 
\eqref{eq_ducc_unitary}, that performs an arbitrary fermionic $n$-tuple particle--hole excitation, 
$\exp(\theta \kappa_{p_1 \ldots p_n}^{p_{n+1} \ldots p_{2n}})$. It is straightforward to 
show that
\begin{equation}\label{eq_jw_ntuple}
	\begin{split}
		e^{\theta \kappa_{p_1 \ldots p_n}^{p_{n+1} \ldots p_{2n}}}
		&\xrightarrow{\text{JW}}
		e^{i\frac{\theta}{2^{2n - 1}} \left(\sum_{l = 1}^{2^{2(n-1)}} P_l
			- \sum_{m = 1}^{2^{2(n-1)}} P_m \right)} \\
		&\xrightarrow{\phantom{\text{JW}}}
		\prod_{l = 1}^{2^{2(n - 1)}} e^{i\frac{\theta}{2^{2n - 1}} P_l}
		\prod_{m = 1}^{2^{2(n - 1)}} e^{-i\frac{\theta}{2^{2n - 1}} P_m},
	\end{split}
\end{equation}
where the $P_i$'s denote Pauli strings. In the last step, we utilized the fact 
that Pauli strings originating from the same second-quantized operator commute \cite{Romero2019}. 
As depicted in Fig.\ \ref{figure1}, the standard quantum circuit 
implementing the product of unitaries appearing in Eq.\ \eqref{eq_jw_ntuple} 
involves a series of single-qubit gates and two-qubit 
CNOTs. The mathematical expressions for the single- and two-qubit gate counts are provided in Table 
\ref{table_analytic_counts}. As shown 
in Table \ref{table_analytic_counts}, the number of CNOT gates scales exponentially with the 
excitation rank $n$, as $\alpha 2^{2n}$, and, to make matters worse, the prefactor $\alpha$ depends 
on the indices of the spinorbitals involved in the excitation process (a manifestation of the 
non-local nature of fermionic sign in the JW encoding).

\begin{turnpage}
	\begin{table*}[t]
		\caption{\label{table_analytic_counts}
			Comparison of the numbers of single-qubit gates and CNOTs involved in the various 
			quantum 
			circuits implementing anti-Hermitian $n$-tuple fermionic and qubit particle--hole 
			excitations, 
			$\kappa_{p_1
				\ldots p_n}^{p_{n+1} \ldots p_{2n}}$, examined in this work.}
		\begin{tabular*}{\linewidth}{@{\extracolsep{\fill}} l c c c c}
			\toprule
			\multirow{2}{*}{Quantum Circuit} & \multicolumn{2}{c}{Analytic Count:
				$\kappa_{p_1 \ldots p_n}^{p_{n+1} \ldots p_{2n}}$} & 
			\multicolumn{2}{c}{Example: $\kappa_{2\;1\;\:\;5}^{8\;9\;11}$} \\
			\cmidrule(lr){2-3} \cmidrule(lr){4-5} 
			& SQG\protect\footnotemark[1] & CNOT & SQG\protect\footnotemark[1] & CNOT \\
			\midrule
			Standard Fermionic\protect\footnotemark[2] & $\displaystyle (4n + 1) 2^{2n - 1}$ & 
			$\displaystyle \left( \sum_{i = 
				0}^{2n - 1} {( - 1)}^i p_{2n - i} + n - 1 \right){2^{2n}}$ & 416 & 512 \\
			Standard Qubit\protect\footnotemark[3] & $\displaystyle (4n + 1) 2^{2n - 1}$ & 
			$\displaystyle (2n - 1)2^{2n}$ & 
			416 & 320 \\
			FEB\protect\footnotemark[4] & $\displaystyle 2^{2n - 1}$ & $\displaystyle 2^{2n - 1} + 
			4n 
			- 2 +
			2\left(\sum_{p_{2i} - p_{2i-1} > 2} (p_{2i} - p_{2i-1} - 2) +
			\sum_{p_{2i} - p_{2i-1} > 1} 1 -
			\left[ \sum_{p_{2i} - p_{2i-1} > 1} 1 > 0\right] \right)$ & 
			32 & 46 
			\\
			QEB\protect\footnotemark[5] & $\displaystyle 2^{2n - 1}$ & $\displaystyle 2^{2n - 1} + 
			4n 
			- 2$ & 32 & 42 \\
			\bottomrule
		\end{tabular*}
		\footnotetext[1]{\setlength{\baselineskip}{1em}
			Single-qubit gate.}
		\footnotetext[2]{\setlength{\baselineskip}{1em}
			The quantum circuit is defined in Fig.\ \ref{figure1}.}
		\footnotetext[3]{\setlength{\baselineskip}{1em}
			The quantum circuit is defined in Fig.\ \ref{figure2}.}
		\footnotetext[4]{\setlength{\baselineskip}{1em}
			The quantum circuit is defined in Fig.\ \ref{figure5}. By taking advantage of certain 
			circuit 
			identities \cite{Yordanov2020}, the reported CNOT counts can be reduced by one. 
			However, doing 
			so would introduce a number of single-qubit gates that scales linearly with the 
			excitation 
			rank. Finally, the CNOT counts do not include the two CZ gates shown in Fig.\ 
			\ref{figure5}.}
		\footnotetext[5]{\setlength{\baselineskip}{1em}
			The quantum circuit is defined in Fig.\ \ref{figure4}. By taking advantage of certain 
			circuit 
			identities \cite{Yordanov2020}, the reported CNOT counts can be reduced by one. 
			However, doing 
			so would introduce a number of single-qubit gates that scales exponentially with the 
			excitation 
			rank.}
		
	\end{table*}
\end{turnpage}

\begin{figure*}[t]
	\centering
	\includegraphics[width=\textwidth]{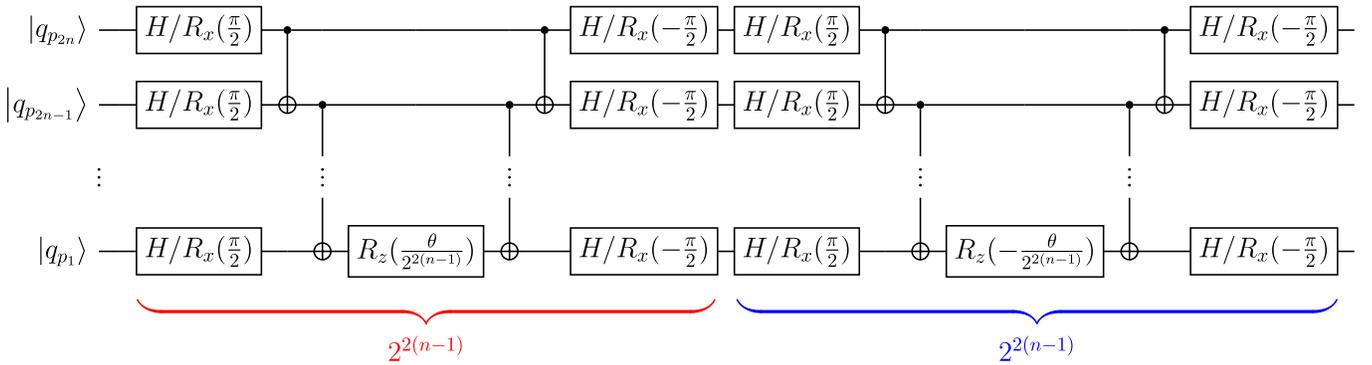}
	\caption{\label{figure2}
		Standard quantum circuit performing a qubit $n$-tuple particle--hole excitation,
		$\exp(\theta Q_{p_1 \ldots p_n}^{p_{n+1} \ldots p_{2n}})$. The vertical dotted lines denote 
		CNOT ``staircases.''
	}
\end{figure*}

One way to reduce the number of CNOTs is by replacing fermionic excitations $\kappa_{p_1 \ldots 
p_n}^{p_{n+1} \ldots p_{2n}}$ by their qubit counterparts $Q_{p_1 \ldots p_n}^{p_{n+1} \ldots 
p_{2n}} \equiv Q^{p_{n+1}} \cdots Q^{p_{2n}} Q_{p_n} \cdots Q_{p_1} - Q^{p_1} \cdots Q^{p_n} 
Q_{p_{2n}} \cdots Q_{p_{n+1}}$, which practically translates to removing the strings of $Z$ gates 
encoding the 
fermionic sign in Eqs.\ \eqref{eq_jw_create} and \eqref{eq_jw_annihilate}.
The standard quantum circuit performing an arbitrary qubit $n$-tuple 
particle--hole excitation is shown in Fig.\ \ref{figure2} and the corresponding counts of 
single-qubit gates and CNOTs are given in Table \ref{table_analytic_counts}. A quick inspection of 
Fig.\ \ref{figure2} immediately reveals the local nature of qubit excitations since all quantum 
gates act exclusively on the qubits involved in the excitation process. Although the CNOT gate 
count is still characterized by the same exponential scaling as in the fermionic case, the 
replacement of fermionic excitations by their qubit counterparts 
results in quantum circuits with substantially reduced CNOT counts. For example, the standard 
quantum circuit performing the qubit triple excitation 
$Q_{2\;1\;\:\;5}^{8\;9\;11}$ contains about 38\% less CNOTs than its 
fermionic cousin. 
This is a consequence of the fact that the number of CNOT gates depends only on the excitation 
rank, \foreign{e.g.}, single, double, and triple qubit excitations will give rise to quantum 
circuits containing 4, 48, and 320 CNOT gates, respectively, independently of the indices involved 
in the excitation processes.

At this point, it is worth mentioning that the reduction in the CNOT count 
associated with qubit excitations might come at the cost of possibly sacrificing the 
proper sign structure of the resulting state. This can be illustrated, for example, by comparing 
the actions of the fermionic and qubit creation operators, \foreign{i.e.},
\begin{equation}
	a^p \ket*{\ldots n_p \ldots} = (1-n_p)(-1)^{\sum_{r=0}^{p-1} n_r} \ket*{\ldots 
		1_p  \ldots},
\end{equation}
\begin{equation}
	Q^p \ket*{\ldots q_p  \ldots} = (1-q_p) \ket*{\ldots 1_p  \ldots}.
\end{equation}
Thus, despite the fact that a wavefunction generated by a qubit approach will have the proper 
fermionic anti-symmetry, which is inherent to the employed many-electron basis of Slater 
determinants \cite{Mazziotti2021}, obtaining the sign structure, up to a phase, of the desired 
state might not be trivial. This implies that qubit schemes might be potentially characterized by a 
slower convergence to the FCI solution compared to their fermionic counterparts.

\subsection{CNOT-Efficient Single and Double Excitations}
\label{sec_QEB_FEB_SD}

\begin{figure}[h]
	\centering
	\includegraphics[width=\linewidth]{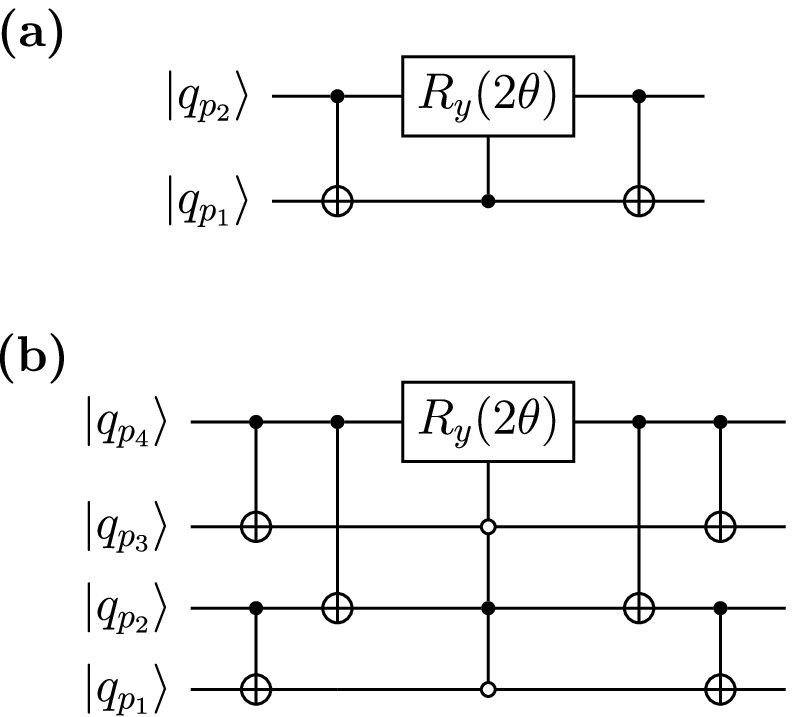}
	\caption{\label{figure3}
		QEB quantum circuits performing (a) single and (b) double particle--hole excitations,
		$\exp(\theta Q_{p_1}^{p_2})$ and $\exp(\theta Q_{p_1 p_2}^{p_3 p_4})$, 
		respectively. The open circles denote anti-control qubits.
	}
\end{figure}

The standard quantum circuits 
performing qubit excitations are not the most economical in terms of CNOT gates. In designing 
CNOT-efficient quantum 
circuits for qubit single and double excitations, Yordanov \foreign{et al.}\ chose a different 
route \cite{Yordanov2020}. Instead of expressing the qubit creation and annihilation operators in 
terms of Pauli gates, they examined the action of the entire unitaries performing qubit single and 
double excitations, namely, $\exp(\theta Q^{p_2}_{p_1})$ and $\exp(\theta 
Q^{p_3 p_4}_{p_1 p_2})$, respectively, on an arbitrary multi-qubit basis state. Based on 
the observation that these unitaries continuously exchange the states $\ket*{1_{p_1} 0_{p_2}}$ and 
$\ket*{0_{p_1} 1_{p_2}}$ in the case of singles and $\ket*{1_{p_1} 1_{p_2} 0_{p_3} 0_{p_4}}$ and 
$\ket*{0_{p_1} 0_{p_2} 1_{p_3} 1_{p_4}}$ for doubles 
while leaving any other basis state unchanged, Yordanov \foreign{et al.}\ showed that such 
controlled exchanges can be performed with the QEB quantum circuits depicted in Fig.\ 
\ref{figure3}. After decomposing the multi-qubit-controlled $R_y 
(2\theta)$ gate in terms of CNOTs and single-qubit $R_y$ rotations 
\cite{Barenco1995,Yordanov2019,Yordanov2020}, the final quantum circuits 
contain 4 and 14 CNOTs for qubit single and double excitations, respectively. By taking advantage 
of circuit identities, the CNOT counts can be decreased to 2 (singles) and 13 (doubles). 
Thus, the replacement of the standard quantum circuits performing qubit single and double 
excitations by their QEB counterparts results in a drastic reduction in the number of CNOTs, 
namely, 50\% for singles and about 73\% in the case of doubles.

Yordanov \foreign{et al.}\ also adapted their QEB quantum circuits for singles and doubles to 
fermionic 
excitations by restoring the fermionic sign information. This was accomplished by ``sandwiching'' 
the quantum circuits of Fig.\ \ref{figure3} between two CNOT ``staircases'' and two controlled Z 
(CZ) gates \cite{Yordanov2020}. The resulting CNOT-efficient FEB quantum circuits required only two 
CNOT ``staircases'' instead of 4 in the case of singles and 16 for doubles.

\subsection{CNOT-Efficient Excitations of Arbitrary Many-Body Rank}
\label{sec_QEB_FEB_n}

\begin{figure}[h]
	\centering
	\includegraphics[width=\linewidth]{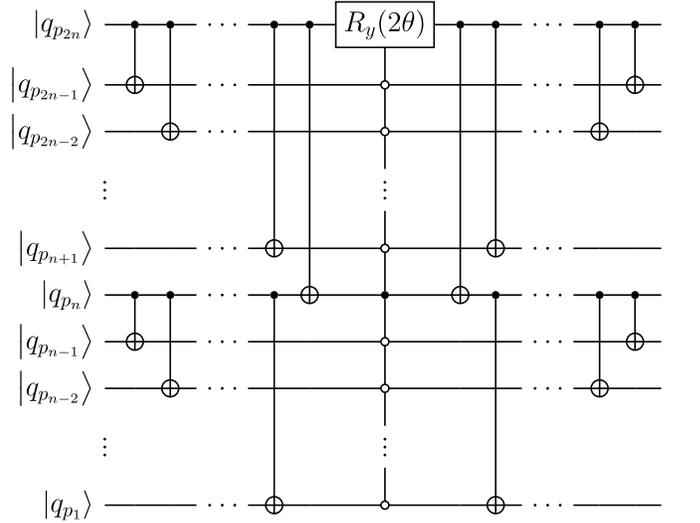}
	\caption{\label{figure4}
		QEB quantum circuit performing a qubit $n$-tuple particle--hole excitation,
		$\exp(\theta Q_{p_1 \ldots p_n}^{p_{n+1} \ldots p_{2n}})$.
	}
\end{figure}

Although Yordanov states that this novel approach to constructing CNOT-efficient quantum circuits 
can be extended to higher-order excitations \cite{Yordanov_thesis}, to the best of our knowledge 
the pertinent analysis has not been carried out and the corresponding optimal circuits have not 
been implemented. To generalize the CNOT-efficient quantum circuits shown in Fig.\ \ref{figure3} to 
an arbitrary excitation rank, we follow a procedure similar to the one of Yordanov \foreign{et 
al.}\ \cite{Yordanov2020}. We begin by examining the action of a unitary 
performing a qubit $n$-tuple excitation on a generic multi-qubit basis state,
\begin{equation}\label{eq_nqubit_general}
	e^{\theta Q^{p_{n+1} \ldots p_{2n}}_{p_1 \ldots p_n}}
	\ket*{q_{p_1}\ldots q_{p_n} q_{p_{n+1}} \ldots q_{p_{2n}}}.
\end{equation}
In principle, depending on the state of the qubits involved in the excitation process, one would 
need to examine $2^{2n}$ distinct cases. Nevertheless, by Taylor expanding the exponential, it is 
straightforward to show that the unitary will leave the $\ket*{q_{p_1}\ldots q_{p_n} q_{p_{n+1}} 
	\ldots q_{p_{2n}}}$ basis state unchanged unless $q_{p_1} = \cdots = q_{p_n} \neq q_{p_{n+1}} = 
	\cdots =
q_{p_{2n}}$, in which case
\begin{equation}\label{eq_nqubit_1_0}
	\begin{split}
		e^{\theta Q^{p_{n+1} \ldots p_{2n}}_{p_1 \ldots p_n}}
		\ket*{1_{p_1}\ldots 1_{p_n} 0_{p_{n+1}} \ldots 0_{p_{2n}}} \\
		= \quad \cos(\theta) \ket*{1_{p_1}\ldots 1_{p_n} 0_{p_{n+1}} \ldots 0_{p_{2n}}} \\
		+ \sin(\theta) \ket*{0_{p_1}\ldots 0_{p_n} 1_{p_{n+1}} \ldots 1_{p_{2n}}}
	\end{split}
\end{equation}
and
\begin{equation}\label{eq_nqubit_0_1}
	\begin{split}
		e^{\theta Q^{p_{n+1} \ldots p_{2n}}_{p_1 \ldots p_n}}
		\ket*{0_{p_1}\ldots 0_{p_n} 1_{p_{n+1}} \ldots 1_{p_{2n}}} \\
		= \quad -\sin(\theta) \ket*{1_{p_1}\ldots 1_{p_n} 0_{p_{n+1}} \ldots 0_{p_{2n}}} \\
		+ \cos(\theta) \ket*{0_{p_1}\ldots 0_{p_n} 1_{p_{n+1}} \ldots 1_{p_{2n}}}
	\end{split}.
\end{equation}
Thus, similarly to single and double qubit excitations, the $\exp(\theta 
Q^{p_{n+1} \ldots p_{2n}}_{p_1 \ldots p_n})$ unitary continuously exchanges 
the $\ket*{1_{p_1}\ldots 1_{p_n} 0_{p_{n+1}} \ldots 0_{p_{2n}}}$ and $\ket*{0_{p_1}\ldots 0_{p_n} 
	1_{p_{n+1}} \ldots 1_{p_{2n}}}$ basis states while acting as the identity otherwise. The 
	resemblance 
of Eqs.\ \eqref{eq_nqubit_1_0} and \eqref{eq_nqubit_0_1} to the action of the single-qubit 
$R_y(2\theta)$ gate on the $\ket*{0}$ and $\ket*{1}$ states, respectively, allows us to construct 
the 
CNOT-efficient QEB quantum circuit shown in Fig.\ \ref{figure4}. The CNOT gates to the left of the 
multi-qubit-controlled $R_y(2\theta)$ gate ensure that the circuit will only act non-trivially on 
the two aforementioned basis states. This can be seen as follows. The cascade of $n - 1$ CNOT gates 
between $q_{p_n}$ and the remaining $q_{p_m}, m = 1, \ldots, n - 1$, qubits checks whether each of 
the $q_{p_m}$ qubits is in the same quantum state as $q_{p_n}$; the action of the CNOT gate will 
result in qubit $q_{p_m}$ being in the $\ket*{0}$ state if $q_{p_n}$ and $q_{p_m}$ are in the same 
state, otherwise $q_{p_m} = \ket*{1}$. The cascade of $n - 1$ CNOTs between $q_{p_{2n}}$ 
and the remaining $q_{p_m}, m = n + 1, \ldots, 2n - 1$, qubits acts in a similar manner.
The single CNOT gate entangling qubits $q_{p_{2n}}$ and $q_{p_n}$ checks whether these two qubits 
are in the same quantum state. Consequently, the multi-qubit-controlled $R_y(2\theta)$ gate will 
only be applied if the following three conditions are satisfied: $q_{p_1} = \cdots = q_{p_n}$, 
$q_{p_{n+1}} = \cdots = q_{p_{2n}}$, and $q_{p_n} \neq q_{2n}$. This analysis justifies the 
choice of control \foreign{vs} anti-control qubits in the multi-qubit-controlled $R_y(2\theta)$ 
gate. If the multi-qubit-controlled $R_y(2\theta)$ gate is not applied, then the CNOT gates to its 
right will pairwise cancel with the CNOT gates to its left, leaving the system in the same quantum 
state. Otherwise, the CNOT gates to the right of the multi-qubit-controlled $R_y(2\theta)$ gate 
complete the desired transformation.

\begin{figure*}[t]
	\centering
	\includegraphics[scale=0.75]{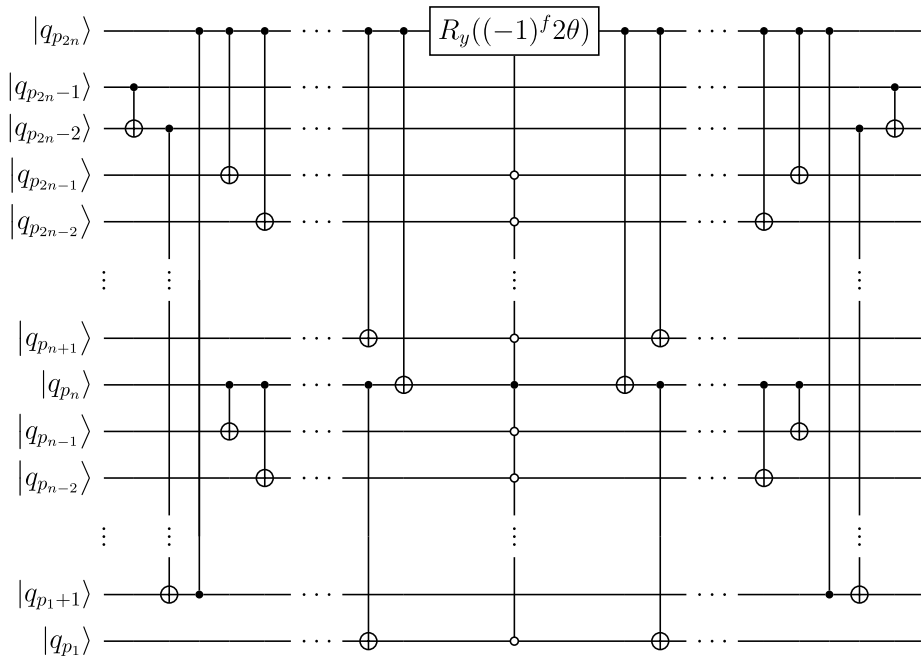}
	\caption{\label{figure5}
		FEB quantum circuit performing a fermionic $n$-tuple particle--hole excitation,
		$\exp(\theta \kappa_{p_1 \ldots p_n}^{p_{n+1} \ldots p_{2n}})$. The parameter
		$f$ controlling the sign of the rotation angle depends on the excitation rank
		$n$ as follows: $f = 0$ for $n = 1, 4, 5, 8, 9, \ldots$ and $f = 1$ for $n =
		2, 3, 6, 7, \ldots$. 
	}
\end{figure*}

As demonstrated in Sec.\ SI of the Supplemental Material, the 
decomposition of 
the multi-qubit-controlled $R_y(2 \theta)$ gate introduces an exponential number of single-qubit 
gates and CNOTs, thus dominating the pertinent total counts reported in Table 
\ref{table_analytic_counts}. Nevertheless, the QEB quantum circuit offers a dramatic reduction in 
both the numbers of single-qubit gates and CNOTs when compared to those characterizing its 
traditional counterpart. Furthermore, the reductions become more impressive as the excitation rank 
becomes larger. For example, the QEB implementations of qubit triple, quadruple, pentuple, and 
hextuple excitations require about 92\%, 94\%, 95\%, and 96\% less single-qubit gates and 87\%, 
92\%, 94\%, and 95\% fewer CNOTs, respectively, when compared to the corresponding standard qubit 
quantum circuits. At this point, it is worth mentioning that, similarly to the QEB singles and 
doubles case, one can utilize circuit identities to reduce the number of CNOTs by 1 when 
decomposing the multi-qubit-controlled $R_y (2 \theta)$ gate. However, as outlined in Sec.\ SII in the 
Supplemental Material, these circuit identities introduce a number of single-qubit gates that 
scales linearly with the excitation rank, essentially outweighing the benefits of reducing the 
operational noise by removing a single CNOT, especially so for higher-rank excitations. 
Consequently, our implementation produces QEB quantum circuits with the single-qubit and CNOT gate 
counts shown in Table \ref{table_analytic_counts}, which we believe offer the best balance between 
CNOTs and overall gate errors.

The FEB quantum circuit performing a CNOT-efficient fermionic $n$-tuple 
excitation is shown in Fig.\ \ref{figure5}. In addition to the two extra CNOT ``staircases'' and CZ 
gates considered by Yordanov \foreign{et al.}, a close inspection of Fig.\ \ref{figure5} reveals a 
further 
modification with respect to the QEB quantum circuit of Fig.\ \ref{figure4}. Due to the fact that 
fermions and qubits do not obey the same algebras \cite{Wu2002}, we found that the angle of the 
multi-qubit-controlled $R_y$ gate contains a sign factor that depends on the excitation rank $n$ in 
a rather non-intuitive way, namely, $+1$ for $n = 1, 4, 5, 8, 9,\ldots$ and $-1$ in the case of $n 
= 2, 3, 6, 7, \ldots$.

As shown in Table \ref{table_analytic_counts}, the use of an FEB quantum circuit instead of its 
standard counterpart results in a dramatic decrease in the number of CNOT gates. 
For example, the FEB quantum circuit performing the $\kappa_{2\;1\;\:\;5}^{8\;9\;11}$ fermionic 
triple excitation contains 91\% less CNOTs than its standard analog. A quick comparison of Figs.\ 
\ref{figure1} and \ref{figure5} reveals that the impressive decrease in the number of CNOTs can be 
partly attributed to the fact that the standard circuit performing fermionic excitations contains a 
number of CNOT ``staircases'' that scales exponentially 
with the excitation rank $n$ while its FEB analog requires only two ``staircases'', independently 
of $n$.
An FEB quantum circuit typically contains a number of CNOT gates 
that is greater than that of its QEB counterpart. Nevertheless, the disparity between the FEB and 
QEB CNOT counts depends on two factors, namely, the excitation rank $n$ and the indices involved in 
the excitation process. In general, for lower-rank excitations, such as singles and doubles, the 
two extra CNOT ``staircases'' are expected to be the primary source of CNOTs in an FEB circuit, 
maximizing the difference between the FEB and QEB CNOT counts. As the excitation rank increases, 
the FEB CNOT count is dominated by the multi-qubit-controlled $R_y$ gate, bridging the gap with 
respect to its QEB analog. Furthermore, in the case of consecutive excitation 
indices, the FEB and QEB quantum circuits will contain the same number of CNOT gates.

\section{Results and Discussion}
\label{sec_results}

The discussion of our numerical results is divided into three parts. We begin in Sec.\ 
\ref{sec_standard_vs_efficient} by examining the savings in terms of CNOT gates offered by the FEB- 
and QEB-SPQE approaches introduced in this work compared to their standard fermionic SPQE and qubit 
qSPQE analogs. Subsequently, in Sec.\ \ref{sec_feb_vs_qeb_spqe}, we compare the performance of the 
FEB-SPQE scheme with its QEB counterpart in terms of four different metrics, namely, errors with 
respect to FCI, number of parameters in the ansatz, CNOT counts, and number of residual element 
evaluations. We conclude the discussion of our results in Sec.\ \ref{sec_feb-spqe_vs_qeb-adapt} 
with a comparison of the most CNOT-efficient SPQE and ADAPT-VQE variants considered in this study. 

All SPQE computations discussed in this section employed a selection threshold of $\Omega = 
\SI{e-2}{\textit{E}_h}$ (see Appendix A for the algorithmic details of the recently proposed SPQE 
scheme). The ADAPT-VQE simulations reported here used a selection criterion of 
\SI{e-3}{\textit{E}_h}. Additional values of the SPQE and ADAPT-VQE selection thresholds have been 
considered in the Supplemental Material. The remaining computational details can be found in 
Appendix B.

\subsection{Standard \foreign{vs} CNOT-Efficient SPQE}
\label{sec_standard_vs_efficient}

\begin{figure}[t]
	\centering
	\includegraphics[width=\linewidth]{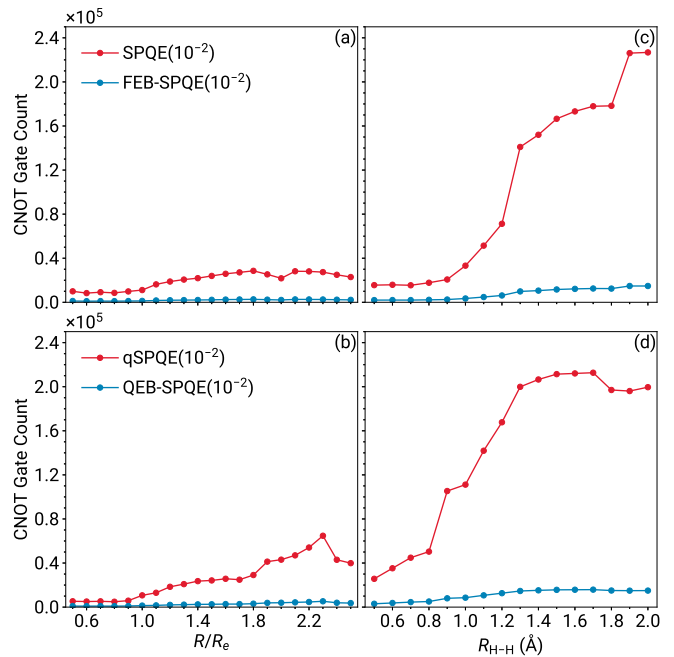}
	\caption{\label{figure6}
		Total CNOT gate counts characterizing the fermionic SPQE and FEB-\spqe{2} [(a) and (c)] 
		and 
		qubit qSPQE and QEB-\spqe{2} [(b) and (d)] ansatz unitaries for the symmetric 
		dissociations 
		of the linear $\text{BeH}_2$ (left column) and $\text{H}_6$ (right column) systems as 
		described by the STO-6G basis.
	}
\end{figure}

The numbers of CNOT gates characterizing the various standard and CNOT-efficient \spqe{2} 
computations of the symmetric dissociations of the linear $\text{BeH}_2$/STO-6G and 
$\text{H}_6$/STO-6G systems are shown in Fig.\ \ref{figure6}. 
A quick inspection of Fig.\ \ref{figure6} reveals that the reduction in the CNOT count offered by 
the FEB- and QEB-\spqe{2} approaches compared to their standard \spqe{2} and q\spqe{2} 
counterparts is significant. Indeed, the replacement of the conventional quantum circuits 
representing 
$n$-tuple fermionic (Fig.\ \ref{figure1}) and qubit (Fig.\ \ref{figure2}) excitations with their 
FEB (Fig.\ \ref{figure5}) and QEB (Fig.\ \ref{figure4}) variants results in about 6--12 times 
less CNOTs in the case of $\text{BeH}_2$ and 8--15 times for $\text{H}_6$.
This is true not only for the equilibrium region of the potential energy curves (PECs), 
characterized by weaker many-electron correlation effects, but also for the stretched 
geometries, where non-dynamical correlations become prevalent. What is even more intriguing 
is the fact that as the strength of the correlation effects increases, the savings in the number 
of CNOTs become more substantial. This can be understood in the following manner. As one 
departs from the weakly correlated regime, the $n$-body excitation operators with $n > 2$ become 
important 
enough to satisfy the SPQE selection criterion. At the same time, the efficacy of the FEB and QEB 
quantum circuits to reduce the number of CNOTs is an increasing function of the excitation rank 
$n$, giving rise to the observed behavior (see Sec.\ \ref{sec_QEB_FEB_n}). 
Although here we focused on the symmetric dissociation of the linear $\text{BeH}_2$ and 
$\text{H}_6$ systems, similar 
observations can be made in the case of the insertion of Be into $\text{H}_2$ and the symmetric 
dissociation of the $\text{H}_6$ ring (see 
Figs.\ S13 and S14 in the Supplemental Material).


The above discussion demonstrates that both the FEB- and QEB-\spqe{2} schemes give rise to 
quantum circuits containing a remarkably smaller number of CNOT gates when compared to those 
obtained with their conventional counterparts. More importantly, the savings in terms of CNOTs are 
anticipated to be even more pronounced in situations involving stronger many-electron correlation 
effects. Furthermore, as illustrated in the Supplemental Material, similarly impressive reductions 
in the CNOT counts are also observed when one employs the less tight importance criterion of 
$\Omega = \SI{e-1}{\textit{E}_h}$ (see Figs.\ S9--S12). Next, we turn our attention to the 
comparison 
of the FEB and QEB flavors of SPQE among themselves.

\subsection{FEB-SPQE \foreign{vs} QEB-SPQE}
\label{sec_feb_vs_qeb_spqe}

The purpose of this section is to determine which of the two CNOT-efficient variants of SPQE offers 
the best balance between accuracy and required computational resources. To gauge the accuracy of 
the fermionic FEB-SPQE approach and its qubit QEB analog, we compare the resulting energetics with 
those obtained from FCI. In estimating the overall 
computational costs of the FEB- and QEB-SPQE schemes, we consider three metrics, namely, the number 
of operators involved in the converged unitary, the number of CNOT gates contained in the quantum 
circuit representing the unitary, and the total number of residual element evaluations 
characterizing the entire SPQE simulation.

\begin{figure}[h]
	\centering
	\includegraphics[width=\linewidth]{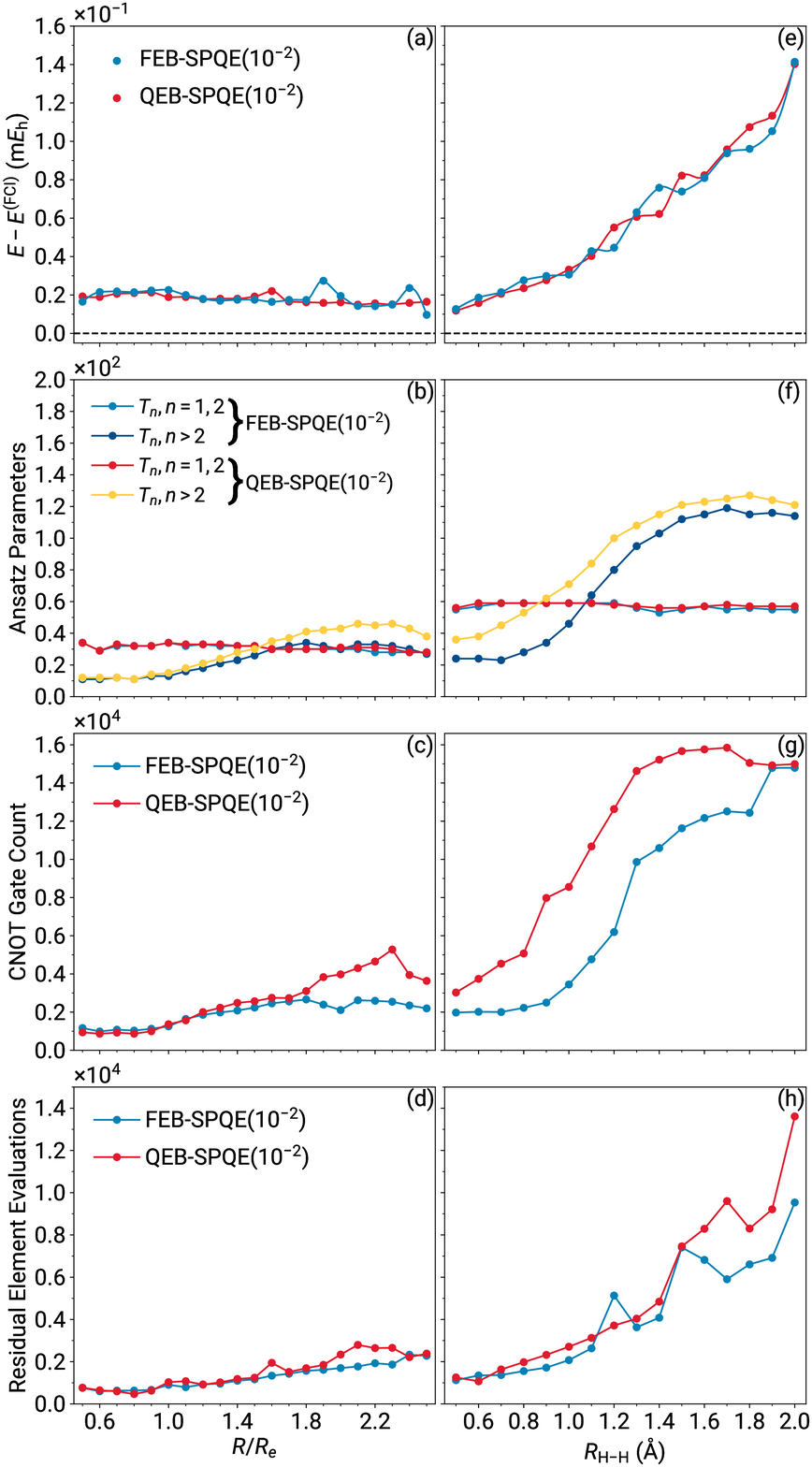}
	\caption{\label{figure7}
		Errors relative to FCI [(a) and (e)], ansatz parameters [(b) and (f)], CNOT gate counts 
		[(c) and (g)], and residual element evaluations [(d) and (h)] characterizing the FEB- and 
		QEB-\spqe{2} simulations of the symmetric dissociations of the linear
		$\text{BeH}_2$ (left column) and $\text{H}_6$ (right column) systems as described by the 
		STO-6G basis.
	}
\end{figure}

In Fig.\ \ref{figure7}, we examine how the FEB- and QEB-\spqe{2} schemes perform when applied to 
the symmetric dissociations of the linear $\text{BeH}_2$ and $\text{H}_6$ species, 
as described by the STO-6G basis set. We begin our 
comparison with a discussion of the pertinent energetics. As illustrated in panels (a) and (e) of 
Fig.\ \ref{figure7}, the energies resulting from the FEB- and QEB-\spqe{2} simulations are not only 
more or less identical among themselves, but also in excellent agreement with those obtained with 
FCI.
Indeed, both CNOT-efficient flavors of SPQE reproduce the exact, FCI, data 
to within a fraction of a millihartree, independently of the strength of the many-electron 
correlation effects. This is even true for the dissociating $\text{H}_6$ linear chain, a 
prototypical system of strong correlations. In this case, although the errors with respect to FCI 
increase as one approaches the dissociation limit, they typically remain one order of magnitude 
below of what is known as ``chemical accuracy,'' namely, \SI{1}{m\textit{E}_h}. At a first glance, 
it might 
seem surprising that QEB-SPQE, which by construction neglects the proper fermionic signs, 
is as accurate as its fermionic counterpart, especially so in situations 
characterized by substantial non-dynamical correlation effects. This behavior is a consequence of 
the tight cumulative threshold of $\Omega = \SI{e-2}{\textit{E}_h}$, which enables both FEB- and 
QEB-SPQE to 
practically recover the FCI solution. Nevertheless, additional insights in this aspect can be 
gained by examining the number of ansatz parameters.

A quick inspection of panels (b) and (f) of Fig.\ \ref{figure7} reveals that the QEB-\spqe{2} 
method produces less compact ans\"{a}tze when compared to the FEB variant, in particular when 
many-electron correlation effects become significant. To make matters worse, 
although the FEB- and QEB-\spqe{2} unitaries contain roughly the same number of single and 
double excitation operators, QEB-\spqe{2} requires a larger number of $n$-tuple excitations with $n 
> 2$ to reach convergence. The explanation of this behavior lies in the fact that qubit excitation 
operators do not account for proper fermionic signs. Therefore, as far as the number of ansatz 
parameters is concerned, it is anticipated that qubit 
approaches will be characterized by a less rapid convergence to the FCI solution when 
compared to their fermionic analogs. As an extreme example of this slower convergence, in Fig.\
S8 we present the errors relative to FCI characterizing the energies obtained with the fermionic 
FEB-UCCSD, FEB-UCCSDT, and FEB-UCCSDTQ schemes and their qubit counterparts for the 
symmetric dissociation of the $\text{H}_6$/STO-6G linear chain. In this particular case, the 
FEB-UCCSD energetics are, in general, more accurate than 
those obtained with the nominally higher-level QEB-UCCSDT scheme. Despite the dramatic improvement 
arising from the incorporation of quadruples, the QEB-UCCSDTQ approach gives rise 
to a PEC that is only slightly better than that obtained with FEB-UCCSDT. At the same time, the 
fermionic FEB-UCCSDTQ potential can hardly be distinguished from its FCI counterpart. These 
observations highlight the fact that higher--than--two-body excitation operators are needed to 
restore the 
proper fermionic sign structure in a many-fermion wavefunction obtained with a qubit ansatz.

Although FEB-\spqe{2} leads to more compact ans\"{a}tze compared to its QEB variant, it is not 
immediately obvious that it will be more CNOT-efficient. To be precise, the CNOT count depends not 
only on the number of operators incorporated in the ansatz unitary, but also on their identity, 
\foreign{i.e.}, the many-body ranks of the operators and the indices involved in the excitation 
processes. As discussed in Sec.\ \ref{sec_QEB_FEB_n}, the number of CNOT gates contained in a 
quantum 
circuit implementing an $n$-tuple fermionic/qubit excitation scales exponentially with $n$. Since 
QEB-\spqe{2} gives rise to unitaries that typically include a larger number 
of higher--than--two-body excitations than its FEB analog, FEB-\spqe{2} is anticipated to 
be, in general, 
more economical in terms of CNOTs. As illustrated in panels (c) and (g) of Fig.\ \ref{figure7}, 
this is indeed the case, especially in situations involving substantial non-dynamic correlation 
effects. For example, for the dissociating $\text{H}_6$ linear chain, a prototypical 
strongly correlated system, the FEB-\spqe{2} approach requires on average 36\% less CNOTs than its 
QEB counterpart. However, the case of the symmetric dissociation of the linear
$\text{BeH}_2$ species is not as straightforward. If we focus on the geometries where both Be--H 
bonds are 
stretched to about twice their equilibrium distance or more, a similar picture to $\text{H}_6$ 
emerges, with the FEB-\spqe{2} CNOT counts being on average 43\% smaller than the QEB ones. The 
situation reverses in favor of QEB-\spqe{2} when one examines the equilibrium region of the 
$\text{BeH}_2$ potential, which is characterized by weaker many-electron correlation effects. In 
this case, despite the fact that the FEB-\spqe{2} ans\"{a}tze are marginally more compact than 
their QEB counterparts, the QEB-\spqe{2} quantum circuits contain a slightly smaller number of CNOT 
gates compared to those resulting from FEB-\spqe{2}. This behavior can be understood by examining 
the operators included in the FEB- and QEB-\spqe{2} unitaries. For these compressed geometries, 
QEB-\spqe{2} typically incorporates one or two additional triple excitations 
with respect to their FEB analogs. Nevertheless, the number of CNOT 
gates introduced by a couple of qubit triple excitation operators is not enough to outweigh the 
number of CNOTs associated with the CNOT ``staircases'' in the fermionic case. This further 
emphasizes the point that, when comparing FEB and QEB approaches, a more compact ansatz may not 
necessarily translate into a more 
CNOT-efficient quantum circuit. 

As a final test of the performance of the FEB- and QEB-\spqe{2} approaches, we compare the total 
number of residual element evaluations characterizing the various simulations. As demonstrated in 
panels (d) and (h) of Fig.\ \ref{figure7}, FEB-\spqe{2} requires, in general, slightly fewer 
residual element evaluations than its QEB counterpart. This can be rationalized based on the fact 
that the QEB-\spqe{2} ans\"{a}tze typically contain a larger number of parameters.

At this point, it is worth mentioning that the above observations regarding the efficiency of 
the FEB- and QEB-SPQE schemes may depend on the value of the cumulative importance criterion 
$\Omega$. For example, as shown in Fig.\ S15 using the 
symmetric dissociations of the linear $\text{BeH}_2$/STO-6G and $\text{H}_6$/STO-6G systems, 
the more relaxed $\Omega = \SI{e-1}{\textit{E}_h}$ threshold results in the distinct advantage of 
FEB-SPQE 
over QEB-SPQE being all but lost. In this case, the FEB- and QEB-\spqe{1} ans\"{a}tze typically 
contain more or less the same number of parameters. As a result, in general, QEB-\spqe{1} gives 
rise to quantum circuits with fewer CNOT gates compared to their FEB counterparts, the only 
exception being the stretched geometries of $\text{H}_6$. Finally, although both the FEB- and 
QEB-\spqe{1} schemes utilize much fewer computational resources compared to their $\Omega = 
\SI{e-2}{\textit{E}_h}$ analogs, they are plagued by substantial errors with respect to the exact 
energies. 

Based on the above analysis, it is evident that FEB-\spqe{2} with $\Omega = \SI{e-2}{\textit{E}_h}$ 
offers the best overall performance. 
It not only generates the energies that faithfully reproduce those obtained from FCI, 
but it also produces compact ans\"{a}tze 
leading to decreased CNOT counts. Although here we focused on the symmetric 
dissociations of the linear $\text{BeH}_2$ and $\text{H}_6$ systems, more or less similar 
observations can be made 
when one examines the insertion of Be into $\text{H}_2$ and the symmetric dissociation of the 
$\text{H}_6$ ring (see Fig.\ S17 in the Supplemental Material).

\subsection{FEB-SPQE \foreign{vs} QEB-ADAPT-VQE}
\label{sec_feb-spqe_vs_qeb-adapt}

Having established that, among the SPQE flavors examined in this work, FEB-\spqe{2}
with $\Omega = \SI{e-2}{\textit{E}_h}$ offers the best 
balance between accuracy and computational cost, we now proceed to compare its performance 
against that of ADAPT-VQE. In an effort to achieve a fair comparison, we first set 
out to determine which of the FEB and QEB versions of ADAPT-VQE is the most efficient, 
\foreign{i.e.}, giving rise to quantum circuits with fewer CNOT gates while producing energies 
within chemical accuracy. A simple inspection of Figs.\ S18--S25 in the Supplemental Material 
reveals that, among the tested ADAPT-VQE schemes,
only FEB- and QEB-\adaptgsd{3} with a selection threshold of \SI{e-3}{\textit{E}_h}
consistently generate energetics within \SI{1}{\milli \textit{E}_h} absolute error from FCI. At the 
same time, QEB-\adaptgsd{3} yields, in general, unitaries with a few 
hundred CNOT gates less. This is true despite the fact that
QEB-\adaptgsd{3} typically gives rise to less compact ans\"{a}tze than its FEB 
counterpart. Nevertheless, the CNOT gates associated with the additional single and double qubit 
excitation operators are not enough to outweigh those arising from the CNOT ``staircases'' 
in the fermionic case. Similar observations regarding the efficiencies of FEB- and QEB-ADAPT-VQE 
were made by Yordanov \foreign{et al.}\ \cite{Yordanov2021}.
Consequently, in what follows, we compare the performance of the 
FEB-\spqe{2} and QEB-\adaptgsd{3} schemes with selection thresholds of \SI{e-2}{\textit{E}_h} and 
\SI{e-3}{\textit{E}_h}, respectively. For the sake of 
completeness, we also include the corresponding results obtained with the less accurate 
QEB-\adaptsd{3} approach, the best ADAPT-VQE scheme considered in this study using a pool of 
particle--hole singles and doubles.

\begin{figure}[t]
	\centering
	\includegraphics[width=\linewidth]{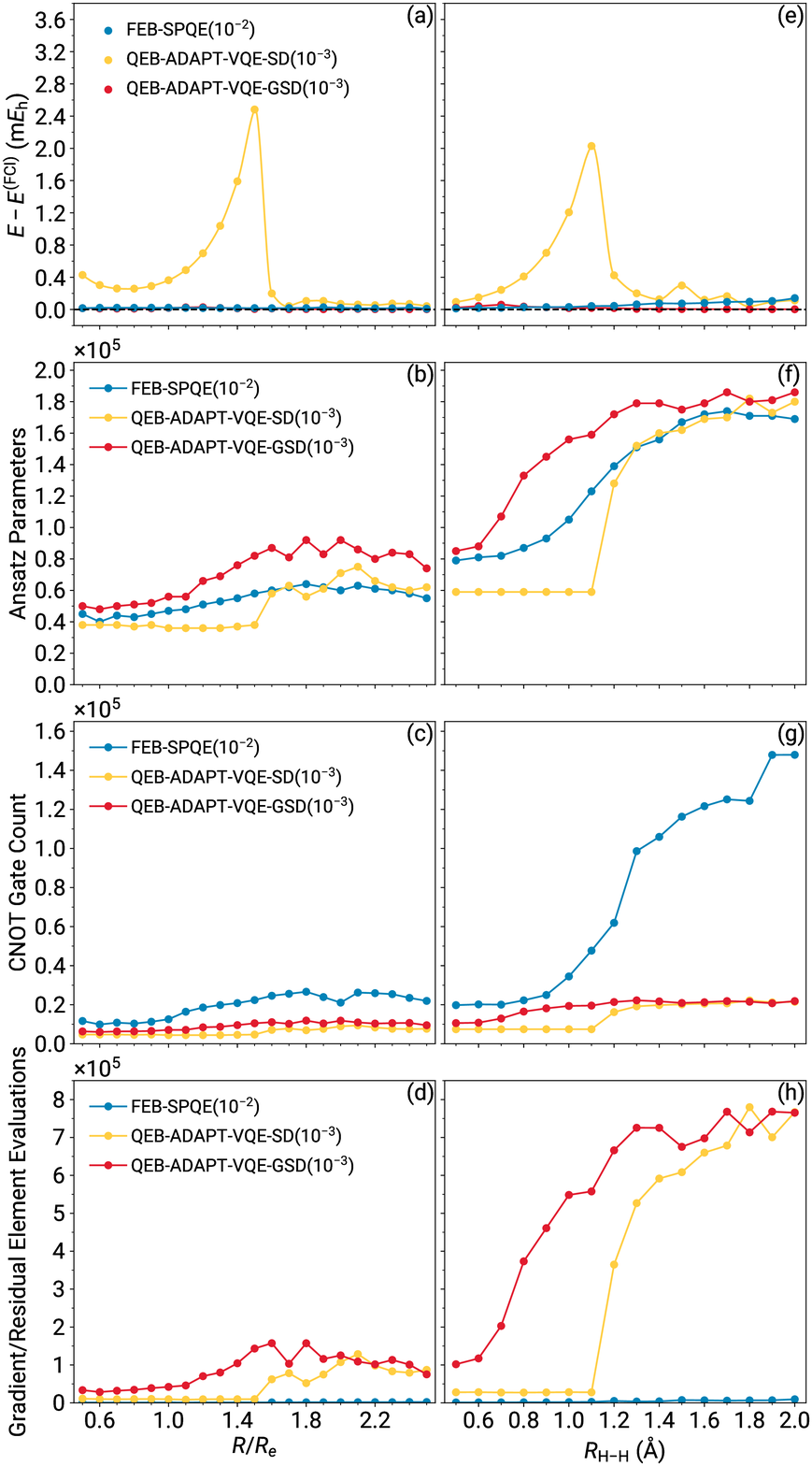}
	\caption{\label{figure8}
		Errors relative to FCI [(a) and (e)], ansatz parameters [(b) and (f)], CNOT gate counts 
		[(c) and (g)], and gradient/residual element evaluations [(d) and (h)] characterizing 
		the FEB-\spqe{2}, QEB-\adaptsd{3}, and QEB-\adaptgsd{3} simulations of the symmetric 
		dissociations of the linear $\text{BeH}_2$ (left column) and $\text{H}_6$ systems (right 
		column) as described by the STO-6G basis.
	}
\end{figure}

In Fig.\ \ref{figure8}, we present the results of our numerical simulations for the
symmetric dissociations of the linear $\text{BeH}_2$/STO-6G and $\text{H}_6$/STO-6G systems as 
obtained with the FEB-\spqe{2}, QEB-\adaptsd{3}, and QEB-\adaptgsd{3} approaches. As was the case 
with the comparison between the FEB and QEB flavors of SPQE, in order to judge which scheme offers 
the best balance between accuracy of the computed energies and required computational resources, we 
rely on four metrics, namely, errors relative to the exact, FCI, energies, compactness of the 
ans\"{a}tze, number of CNOT gates contained in the quantum circuits representing the ansatz 
unitaries, and number of residual [FEB-\spqe{2}] and gradient [QEB-\adaptsd{3}, QEB-\adaptgsd{3}] 
element evaluations. 

We begin our discussion with a comparison of the energetics. A quick inspection of panels (a) and 
(e) of Fig.\ \ref{figure8} immediately reveals that QEB-\adaptsd{3} is the least accurate among the 
three examined schemes. Indeed, the errors relative to FCI characterizing the QEB-\adaptsd{3} 
simulations are much larger than those resulting from QEB-\adaptgsd{3} and FEB-\spqe{2}. To make 
matters worse, QEB-\adaptsd{3} is characterized by maximum errors that exceed 
\SI{2}{m\textit{E}_h}, 
\foreign{i.e.}, more than two times larger than what is known as chemical accuracy, even in the 
less complicated case of $\text{BeH}_2$. Switching to an operator pool of generalized singles and 
doubles dramatically improves the results. In fact, the QEB-\adaptgsd{3} scheme produces PECs for 
the symmetric dissociations of the linear $\text{BeH}_2$/STO-6G and $\text{H}_6$/STO-6G systems 
that can hardly be distinguished from those resulting from the exact, FCI, calculations. The 
FEB-\spqe{2} approach is competitive with QEB-\adaptgsd{3} in 
the weakly correlated regime and in situations involving moderately strong correlations. Indeed, 
the FEB-\spqe{2} and QEB-\adaptgsd{3} energies are more or less identical, differing by just a few 
microhartrees, in the case of the double-bond dissociation of the linear $\text{BeH}_2$ species and 
for the equilibrium region of the $\text{H}_6$ potential. Although as one 
approaches the strong correlation limit in the latter case FEB-\spqe{2} becomes somewhat less 
accurate than QEB\adaptgsd{3}, it remains practically one order of magnitude below what is 
considered chemical accuracy.

We now turn our attention to the numbers of ansatz parameters. As shown in panels (b) and (f) of 
Fig.\ \ref{figure8}, the least accurate QEB-\adaptsd{3} scheme produces, in general, the 
most compact ans\"{a}tze, typically incorporating fewer excitation operators than FEB-\spqe{2} and 
QEB-\adaptgsd{3}. At the other end of the spectrum, QEB-\adaptgsd{3}, which faithfully reproduces 
the exact, FCI, energies, requires the largest number of operators to reach convergence. 
FEB-\spqe{2} always requires fewer, some times much fewer, ansatz parameters than QEB-\adaptgsd{3}, 
but typically more than QEB-\adaptsd{3}. Nevertheless, it is worth mentioning that in situations 
involving significant non-dynamic correlation effects, FEB-\spqe{2} becomes competitive with 
QEB-\adaptsd{3}, even generating the most compact ans\"{a}tze for a few nuclear configurations.

Despite the excellent performance of FEB-\spqe{2} in terms of energetics and ansatz parameters, 
it always generates quantum circuits containing more CNOT gates than both of the examined flavors 
of ADAPT-VQE. As illustrated in panels (c) and (g) of Fig.\ \ref{figure8}, the disparity between 
the FEB-\spqe{2} and QEB-\adaptsd{3} and QEB-\adaptgsd{3} CNOT counts is exacerbated as the 
strength of non-dynamic correlation effects increases. For example, focusing on the strongly 
correlated $\text{H}_6$ linear chain and the largest H--H separation considered in this work, 
namely, $R_\text{H--H} = \SI{2.0}{\AA}$, FEB-\spqe{2} requires about seven times more CNOT gates 
than QEB-\adaptsd{3} or its GSD counterpart.
As might have been anticipated, the source behind the disparity in 
the CNOT counts can be traced to the identities of the excitation operators included in the 
respective operator pools. In the ADAPT-VQE simulations shown in Fig.\ \ref{figure8} we use only 
(generalized) single and double excitations while in the case of SPQE we employ a full operator 
pool, including up to $N$-tuple excitation operators, where $N$ is the number of correlated 
electrons. In light of the fact that the number of CNOT gates introduced by a given $n$-tuple 
excitation operator scales exponentially with the many-body rank $n$, it is not surprising that 
FEB-\spqe{2} requires, in general, many more CNOT gates than QEB-\adaptsd{3} or its GSD variant. 
To put it into perspective, the quantum circuit implementation of the single hextuple excitation in 
$\text{H}_6$/STO-6G, which is incorporated in the FEB-\spqe{2} ansatz already at the $R_\text{H--H} 
= \SI{1.3}{\AA}$ distance between neighboring H atoms, requires essentially the same number of 
CNOTs as that needed for the entire QEB-\adaptsd{3} or QEB-\adaptgsd{3} unitaries.
Nevertheless, it is worth mentioning that, for the same reason, the savings in terms of CNOTs 
afforded by the replacement of the standard fermionic and qubit quantum circuits by their FEB and 
QEB counterparts, respectively, are more impressive in SPQE than ADAPT-VQE. Consequently, by just 
utilizing the CNOT-efficient FEB and QEB quantum circuits instead of their conventional analogs, we 
were able to reduce the disparity in the number of CNOT gates between SPQE and ADAPT-VQE from a 
factor of $\sim 15$ \cite{Stair2021} to a factor of $\sim 7$. 

As a final gauge of the computational resources needed by SPQE and ADAPT-VQE simulations, we 
compare the total numbers of residual [FEB-\spqe{2}] and gradient [QEB-\adaptsd{3}, 
QEB-\adaptgsd{3}] element evaluations. As illustrated in panels (d) and (h) of Fig.\ \ref{figure8}, 
QEB-\adaptsd{3} and its GSD counterpart typically require one to two orders of magnitude more 
gradient element evaluations than the number of residual element measurements in FEB-\spqe{2}. This 
colossal difference can be largely attributed to the use of a cumulative importance criterion and 
the direct inversion of the iterative subspace (DIIS) \cite{Pulay1980,Pulay1982,Scuseria1986} 
accelerator in SPQE. To verify this hypothesis, we performed an additional FEB-\spqe{2} 
computation for the $R_\text{H--H} = \SI{2.0}{\AA}$ geometry of the $\text{H}_6$ linear chain, in 
which a single operator was added to the ansatz per macro-iteration and the DIIS accelerator was 
turned off in the micro-iterations, as is typically done in ADAPT-VQE simulations. Based on this 
numerical experiment, although the number of residual evaluations increased from less than 10,000 
to about 500,000, it still remained about 30\% smaller than the number of gradient evaluations 
in QEB-\adaptsd{3} and QEB-\adaptgsd{3}. At the same time, 
expanding the SPQE unitary one operator at a time yielded a more compact FEB-\spqe{2} ansatz, 
containing 157 parameters instead of 169. As a result, the number of CNOT gates was also decreased 
by about 1,600, further bridging the gap between the SPQE and ADAPT-VQE CNOT counts. Finally, the 
reduction in the number of ansatz parameters was accompanied by a tiny increase, of about 
\SI{30}{\micro \textit{E}_h}, in the error relative to the FCI energy.

Although in Fig.\ \ref{figure8} we focused on the symmetric dissociations of the linear 
$\text{BeH}_2$ and $\text{H}_6$ species, similar remarks can be made for the 
insertion of Be into $\text{H}_2$ and the symmetric dissociation of the 
$\text{H}_6$ ring, considered in the Supplemental Material. Indeed, as illustrated in Fig.\ S26, 
FEB-\spqe{2} is practically as accurate as 
QEB-\adaptgsd{3} while leading to more compact ans\"{a}tze and requiring orders of magnitude less 
residual element evaluations than gradient element measurements in QEB-\adaptgsd{3}. It is also 
interesting to note that, even though FEB-\spqe{2} gives rise to quantum circuits containing, in 
general, a larger number of CNOT gates, as one moves toward the reactants in the 
reaction pathway of $\text{BeH}_2$, FEB-\spqe{2} requires less CNOTs than QEB-\adaptgsd{3} and, 
eventually, QEB-\adaptsd{3}.

The above analysis demonstrates that FEB-\spqe{2} is an attractive alternative to QEB-\adaptgsd{3}, 
as it generates the similar energies, closely reproducing those obtained from the exact, FCI, 
Hamiltonian diagonalizations, while requiring fewer ansatz parameters. Although FEB-\spqe{2} 
typically requires longer circuits, containing as much as 6--7 times more CNOT gates, the runs are 
much shorter due to the fact that FEB-\spqe{2} needs a few orders of magnitude less residual 
element evaluations than gradient element measurements in QEB-\adaptgsd{3}. 

\section{Conclusions}
\label{sec_conclusions}

In this work, we generalized the CNOT-efficient quantum circuits for single and double 
excitations of Ref.\ \cite{Yordanov2020} to operators of arbitrary 
many-body ranks. We demonstrated that, although the FEB/QEB quantum circuits performing 
$n$-tuple fermionic/qubit excitations still contain a number of CNOT gates that 
scales exponentially with the excitation rank $n$, they are much more economical than their 
conventional counterparts. For a given $n$-tuple excitation operator, the FEB scheme will 
typically require more CNOT gates than QEB, needed to encode the proper fermionic sign. 
Nevertheless, we showed that the relative difference between the FEB and QEB CNOT counts is 
decreased as the many-body rank $n$ increases.

Utilizing these CNOT-efficient quantum circuits, we introduced the FEB and QEB variants of SPQE, a 
hybrid quantum--classical approach that relies on a complete operator pool of particle--hole 
excitations to iteratively construct the ansatz. By performing numerical 
simulations for small molecular systems, we demonstrated that FEB- and QEB-SPQE dramatically reduce 
the CNOT counts relative to their standard counterparts, typically requiring about 6--15 times less 
CNOTs. We also found that the remarkable savings in terms of CNOTs offered by the FEB and QEB 
versions of SPQE are even more pronounced in situations involving stronger many-electron 
correlation effects, which is an encouraging behavior.

When comparing the FEB and QEB variants of SPQE among themselves, we found that the 
FEB-SPQE scheme offered the best balance between accuracy and computational resources. 
According to our numerical simulations, while both FEB- and QEB-SPQE produced energies that 
closely followed those resulting from FCI, FEB-SPQE gave rise to more compact ans\"{a}tze whose 
implementation typically required less CNOT gates than QEB-SPQE. This was particularly true at the 
strong correlation regime, with QEB-SPQE requiring up to 2--3 times more CNOTs than its FEB analog. 

Finally, we compared the performance of FEB-SPQE and QEB-ADAPT-VQE, the best SPQE and ADAPT-VQE 
flavors considered in this study. We showed that both FEB-SPQE and QEB-ADAPT-VQE-GSD provide 
energies well within chemical accuracy, with the latter being more accurate in the presence of 
strong correlations. We found that, although FEB-SPQE provided more compact ans\"{a}tze than 
QEB-ADAPT-VQE-GSD and required orders of magnitude less residual element evaluations than 
gradient element measurements in QEB-ADAPT-VQE-GSD, it required many more CNOT gates. Nevertheless, 
we demonstrated that the 
replacement of the traditional quantum circuits by their CNOT-efficient analogs resulted in a 
dramatic decrease in the disparity between the SPQE and ADAPT-VQE CNOT counts.

The colossal decrease in the CNOT counts afforded by the FEB variant of SPQE is certainly 
remarkable. Nevertheless, the experimental realization of SPQE on NISQ hardware requires even fewer 
numbers of CNOT gates. In the future work, we will explore various avenues, on top of the FEB and 
QEB formalisms, to improve the CNOT-efficiency of SPQE. One such option is to combine FEB-/QEB-SPQE 
with the qubit tapering procedure of Refs.\ \cite{Bravyi2017,Setia2020}, where one takes advantage 
of $\mathbb{Z}_2$ symmetries in the Hamiltonian to eliminate redundant 
qubits from the simulations. An advantage of qubit tapering is that it does not introduce any 
approximations, as was the case with the FEB/QEB representations of standard fermionic/qubit 
quantum circuits. Another possibility is to limit the incorporation of the higher-rank excitation 
operators in the ansatz, since they are the most demanding in terms of CNOTs. This can be 
accomplished, for example, by using a more relaxed cumulative threshold value or adding a penalty 
in the selection criterion that is proportional to the number of CNOT gates of a given operator. To 
still incorporate the physics associated with the missing important higher-rank excitation 
operators, a non-iterative energy correction based on the moments of CC equations 
\cite{Jankowski1991,leszcz,ren1} could be considered. An alternative strategy 
geared towards more CNOT-efficient FEB-/QEB-SPQE variants is to use an entirely different operator 
pool. Following the ADAPT-VQE paradigm, of particular interest in this direction is a pool of 
generalized singles and doubles, whose 
quantum circuit representations contain a substantially smaller number of CNOT gates compared to 
those of $n$-tuple particle--hole excitations in conventional SPQE.


\section*{Acknowledgments}

This work is supported by the U.S.\ Department of Energy under Award No.\ DE-SC0019374 and the NSF 
under Grant No.\ CHE-2038019.


\section*{APPENDIX A: Selected Projective Quantum Eigensolver}
\label{sec_spqe}

\renewcommand{\theequation}{A.\arabic{equation}}
\setcounter{equation}{0}

In this appendix, we summarize the salient features of the SPQE algorithm, which serves as a 
testing ground for illustrating the benefits of the CNOT-efficient FEB/QEB quantum circuits 
discussed in the previous sections. In particular, we focus on applications of SPQE
to electronic structure theory and refer the interested reader to Ref.\ 
\cite{Stair2021} for the additional details.

PQE is a hybrid quantum--classical approach for optimizing the parameters of a trial state, Eq.\ 
\eqref{eq_trial_state}. In typical applications to quantum 
chemistry, the reference state $\ket*{\Phi}$ is a single Slater determinant, usually the HF 
determinant, that can be easily realized on a quantum computer and the unitary operator 
$U(\mathbf{t})$ is, in general, a UCC unitary, Eq.\
\eqref{eq_ducc_unitary}. Similar to conventional CC 
theory, in PQE we obtain the optimum parameters defining the ground-electronic state 
and associated energy of a given system by solving the many-electron Schr\"{o}dinger 
equation projectively, Eq.\ \eqref{eq_pqe_res_con}.
The exact eigenstate $\ket*{\Psi_0}$ and energy $E_0$ can be obtained by employing a full UCC 
unitary and enforcing the residual condition Eq.\ \eqref{eq_pqe_res_con} for all excited 
Slater determinants $\ket*{\Phi_\mu}$ afforded by the one-electron basis. In practice, however, one 
utilizes an 
incomplete operator pool and, thus, Eq.\ \eqref{eq_pqe_res_con} can only be satisfied for the 
subset of Slater determinants $R = \{\ket*{\Phi_\mu} : \kappa_\mu \ket*{\Phi} = \ket*{\Phi_\mu}\}$, 
giving rise to an approximate energy $E_\text{PQE}(\mathbf{t}) > E_0$ [see Eq.\ \eqref{eq_ev}].

A crucial step in the PQE algorithm is the evaluation of the residual elements $r_\mu 
(\mathbf{t})$, whose number equals that of the excitation operators defining a given UCC ansatz. 
Since computing the UCC residuals on a classical machine is intractable, 
we leverage a quantum computer to measure the residual elements. Our group has 
recently demonstrated how to efficiently evaluate the exact residuals $r_\mu (\mathbf{t})$, which 
are off-diagonal matrix elements of the UCC similarity-transformed Hamiltonian $\bar{H} 
(\mathbf{t}) \equiv U^\dagger (\mathbf{t}) H U (\mathbf{t})$, as linear combinations of expectation 
values of $\bar{H} (\mathbf{t})$ with respect to states that can be easily prepared on a quantum 
device \cite{Stair2021}. Indeed, taking advantage of the fact that
\begin{equation}\label{eq_omega}
	\ket*{\Omega_\mu (\theta)} = e^{\theta \kappa_\mu}\ket*{\Phi} = \cos(\theta) \ket*{\Phi} + 
	\sin(\theta) \ket*{\Phi_\mu},
\end{equation}
and that the wavefunction is real, the exact residual element $r_\mu (\mathbf{t})$ can be expressed 
as
\begin{equation}\label{eq_exact_res}
	\begin{split}
	r_\mu (\mathbf{t}) =& \ev*{\bar{H}(\mathbf{t})}{\Omega_\mu (\frac{\pi}{4})}\\
	&-\frac{1}{2} \left( \ev*{\bar{H}(\mathbf{t})}{\Phi_\mu} +
	\ev*{\bar{H}(\mathbf{t})}{\Phi} \right).
	\end{split}
\end{equation}
Since the $\ev*{\bar{H} (\mathbf{t})}{\Phi}$ quantity needs to be computed only once per PQE 
iteration, the computation of an exact residual element requires the evaluation of only two energy 
expectation values, \foreign{i.e.}, it has a comparable cost to the measurement of an exact 
gradient element in VQE via the shift rule \cite{Schuld2019,Kottmann2021}. After measuring on the 
quantum device the residuals corresponding to the operators defining the UCC unitary, a classical 
computer is used to update the optimization parameters,
\begin{equation}
	t_\mu^{(n+1)} = t_\mu^{(n)} + \frac{r_\mu^{(n)}(\mathbf{t})}{\Delta_\mu},
\end{equation}
where the superscripts ``($n$)'' and ``($n+1$)'' denote quantities at iterations $n$ and $n+1$, 
respectively, and $\Delta_\mu \equiv \Delta_{i_1 \ldots i_n}^{a_1 \ldots a_n} = \epsilon_{i_1} + 
\cdots + \epsilon_{i_n} - \epsilon_{a_1} - \cdots - \epsilon_{a_n}$ are M\o{}ller--Plesset 
denominators with $\epsilon_p$ being the HF energy of the $p^\text{th}$ spinorbital (see Ref.\ 
\cite{Stair2021} for the details).

The PQE algorithm described thus far pertains to fixed UCC ans\"{a}tze, meaning UCC unitaries 
defined using a constant number of excitation operators. The most widely used 
procedure for constructing such approaches is to incorporate into the UCC unitary all operators up 
to a 
given excitation rank, giving rise to the UCCSD (UCC with singles and doubles), UCCSDT (UCC with 
singles, doubles, and triples), UCCSDTQ (UCC with singles, doubles, triples, and quadruples), 
\foreign{etc.}\ hierarchy of methods that systematically converges to the exact, FCI, solution. 
Although in the weakly correlated regime the basic UCCSD approach might be enough to obtain 
energetics within chemical accuracy, as one approaches the strong correlation limit the importance 
of higher--than--two-body excitations is anticipated to gradually increase. In 
such situations, higher-level approaches, \foreign{e.g.}, UCCSDT, UCCSDTQ, \foreign{etc.}, need to 
be employed. 
However, due to the inherent rigidity of fixed ans\"{a}tze, one also introduces a number of 
superfluous excitation operators that unnecessarily increase the computational cost.

This issue can be remedied by adopting ans\"{a}tze that are iteratively 
constructed by selecting operators from a given operator pool based on some importance criterion. 
Such schemes have been widely employed in computational chemistry in the form 
of various selected CI approaches \cite{Whitten1969,Bender1969,Huron1973,Buenker1974} and CI/CC 
quantum Monte Carlo methods \cite{Booth2009,Thom2010}. In the realm of quantum computing, the first 
such adaptive approach was ADAPT-VQE \cite{Grimsley2019}, which utilized an operator pool of 
singles and doubles or their generalized version and a selection criterion based on the magnitude 
of the individual gradient elements.
To ensure that the resulting ADAPT-VQE state would eventually converge to the 
exact, FCI, solution, the operator pool was not allowed to ``drain'', meaning that a given operator 
could be selected multiple times. Recently, our group has 
generalized the PQE approach to a flexible, iteratively constructed, ansatz, giving rise to the 
SPQE scheme. In contrast to ADAPT-VQE, the SPQE method employs a full operator pool, 
\foreign{i.e.}, containing up to $N$-tuple particle--hole excitations with $N$ being the number of 
correlated electrons, and an importance criterion based on the residuals. According to the 
Gershgorin circle theorem, the error of the PQE energy with respect to FCI is 
bound by the sum of the absolute value of the residuals corresponding to operators not included in 
the ansatz, $\abs{E_\text{PQE}(\mathbf{t}) - E_0} \le \sum_\nu \abs{r_\nu(\mathbf{t})}$. Thus, by 
incorporating 
in the ansatz the operator corresponding to the residual element with the largest magnitude and 
solving the new set of PQE equations, one reduces the error bound by the largest amount. This is 
the main philosophy behind the SPQE methodology.

In principle, one could use Eq.\ \eqref{eq_exact_res} to compute all residuals 
corresponding to operators not included in the ansatz. However, a much more efficient way is to 
extract these residuals from repeated measurements of a single, suitably prepared, state on a 
quantum device. Ideally, one would prepare and measure the state $\ket*{r(\mathbf{t})} = 
U^\dagger(\mathbf{t}) H U(\mathbf{t}) \ket*{\Phi}$, but this involves applying the non-unitary 
Hamiltonian operator $H$. This can be circumvented by replacing the Hamiltonian by the time 
evolution operator $e^{i \Delta t H}$ and using a small time step $\Delta t$ so that the 
contributions of non-linear terms become negligible,
\begin{equation}
	\begin{split}
		\ket*{\tilde{r}(\mathbf{t})} =& U^\dagger(\mathbf{t}) e^{i \Delta t H} U(\mathbf{t}) 
		\ket*{\Phi} 
		\\
		=& \ket*{\Phi} + i \Delta t U^\dagger(\mathbf{t}) H U(\mathbf{t}) \ket*{\Phi} + 
		\mathcal{O}(\Delta t^2) \\
		=& [1 + i \Delta t E_\text{PQE}(\mathbf{t})] \ket*{\Phi} \\
		&+ i \Delta t \sum_\nu r_\nu(\mathbf{t}) \ket*{\Phi_\nu}+ \mathcal{O}(\Delta t^2).
	\end{split}
\end{equation}
Using the fact that, independently of the fermionic encoding, there is a one-to-one correspondence 
between Slater determinants and elements of the computational basis, one  can estimate the desired 
residuals by performing a sufficiently large number of measurements $M$ of identically prepared 
$\ket*{\tilde{r}}$ states,
\begin{equation}\label{eq_res_sq}
	\abs{r_\nu}^2 \approx \frac{1}{\Delta t^2}\frac{M_\nu}{M},
\end{equation}
where $M_\nu$ is the number of times the state $\ket*{\Phi_\nu}$ was obtained.

After evaluating the residuals corresponding to operators not included in the current ansatz, the 
SPQE algorithm proceeds to the next step, namely, the selection of important missing excitations. 
In contrast to typical ADAPT-VQE applications where 
the ansatz is expanded one operator per macro-iteration, in SPQE we rely on a cumulative importance 
criterion, allowing us to add multiple operators at a time, thus accelerating convergence. To be 
precise, the operators are ordered in ascending order according to $\abs{r_\nu}^2$, Eq.\ 
\eqref{eq_res_sq}, and, starting from the first element, we keep discarding operators until the 
following relation is satisfied:
\begin{equation}\label{eq_cumulative}
	\sum_{\text{ discarded}} \abs{r_\nu}^2 \le \Omega^2,
\end{equation}
where $\Omega^2$ is a user-defined threshold. The remaining set of excitation operators are added 
to the ansatz in order of decreasing $\abs{r_\nu}^2$. By appending to the ansatz a batch of 
operators at a time, SPQE simulations typically converge after a few macro-iterations, dramatically 
reducing the number of residual element evaluations, while still producing compact ans\"{a}tze.

\section*{APPENDIX B: Computational Details}
\label{sec_computational}

\renewcommand{\theequation}{B.\arabic{equation}}
\setcounter{equation}{0}

The CNOT-efficient FEB and QEB quantum circuits performing respectively fermionic and qubit 
excitations of arbitrary order, along with the standard circuits for qubit excitations, have been 
implemented in a local version of the QForte package \cite{qforte}. To gauge the extent to 
which the FEB and QEB quantum circuits reduce the number of CNOT gates compared to their standard 
analogs, in particular when higher-than-double excitations are involved, we performed a series of 
SPQE classical numerical simulations for small molecular systems using 
a full particle--hole excitation operator pool. To ensure the importance of higher--than--two-body 
excitations, we studied molecular processes in which the strength of the many-electron correlation 
effects can be continuously varied from weak to strong. Specifically, we examined the symmetric 
dissociations of the linear $\text{BeH}_2$ and $\text{H}_6$ systems, both treated with the 
minimum STO-6G basis \cite{Hehre1969} and correlating all electrons. In the computations for 
$\text{BeH}_2$, we employed the grid of Be--H internuclear distances $0.5 R_e$, $0.6 R_e$, \ldots, 
$2.5 R_e$, where $R_e = \SI{1.310011}{\AA}$ is the FCI/STO-6G equilibrium distance. The potential 
energy curves characterizing the symmetric dissociation of the $\text{H}_6$ linear chain were 
determined using the following uniform grid of distances between neighboring hydrogen atoms, 
$R_{\text{H--H}}$: 0.5, 0.6, \ldots, \SI{2.0}{\AA}. In the Supplemental Material, we considered two 
additional processes, namely, the insertion of Be into $\text{H}_2$ and the 
symmetric dissociation of the $\text{H}_6$ ring. As was the case with the 
computations of their linear counterparts, we employed the STO-6G basis set and correlated all 
electrons. The reaction pathway defining the insertion of Be into $\text{H}_2$ was constructed as 
follows. We first optimized the geometries of the reactants, $\text{Be} + \text{H}_2$, and the 
product, $\text{BeH}_2$, at the FCI/STO-6G level of theory. Following the strategy of Ref.\ 
\cite{Evangelista2011}, we kept the Be atom fixed at the center of a two-dimensional coordinate 
system and used the optimized structures to define the lines on which the two H atoms move during 
the insertion reaction, namely,
\begin{equation}\label{eq_beh2_c2v}
	y(x) = \pm(0.164183x - 1.310011),
\end{equation}
where the $x$ coordinate takes values between \SI{0}{\AA} (product, $\text{BeH}_2$) and 
\SI{5.746708}{\AA} (reactant, $\text{Be} + \text{H}_2$). Finally, starting from the optimized 
geometry of the product, we sampled the reaction coordinate by selecting 30 points with $x$ 
coordinates given by
$x_n (\mathrm{\AA}) = \frac{5.746708}{29}n$, $n = 0,1,\ldots,29$.
In the case of the dissociation of the $\text{H}_6$ ring, we 
employed the same grid of H--H internuclear separations as in its linear counterpart.

All SPQE simulations performed in this study utilized a full particle--hole excitation 
operator pool, \foreign{i.e.}, containing up to hextuple excitations for both the $\text{BeH}_2$ 
and $\text{H}_6$ systems. We considered two values of the parameter $\Omega$ defining the 
cumulative importance criterion, Eq.\ \eqref{eq_cumulative}, namely, \SI{e-1}{\textit{E}_h} and 
\SI{e-2}{\textit{E}_h} (results for the former case are reported in the Supplemental Material).
All PQE calculations employed a micro-iteration threshold of \SI{e-5}{\textit{E}_h} for the 
residual norm 
$\norm{\mathbf{r}}$ and the DIIS approach was utilized to accelerate convergence. Typically, the 
PQE 
cycles of an SPQE computation converged within 10 to 20 micro-iterations, although there were a few 
exceptions requiring a far greater number. Consequently, in order to maintain a relatively low 
count of residual element evaluations, the maximum number of micro-iterations was set to 50. 

For each of the systems examined in this study, we also carried out ADAPT-VQE simulations using two 
kinds of operator pools, namely, single and double particle--hole excitations and generalized 
singles and doubles. As is typically done in ADAPT-VQE, a single operator was added to the ansatz 
per macro-iteration and the pool was not allowed to ``drain''. For each operator pool, we performed 
two sets of calculations employing the macro-iteration gradient thresholds of 
\SI{e-2}{\textit{E}_h} and \SI{e-3}{\textit{E}_h} (with results for the former being reported in 
the Supplemental Material).
All VQE computations utilized the Broyden--Fletcher--Goldfarb--Shanno (BFGS) optimizer 
\cite{Broyden1970,Fletcher1970,Goldfarb1970,Shanno1970}, as implemented in SCIPY \cite{scipy}, and 
a micro-iteration convergence criterion of \SI{e-5}{\textit{E}_h} for the 
gradient norm $\norm{\mathbf{g}}$. In general, 
the VQE cycles of an ADAPT-VQE simulation required many micro-iterations to achieve convergence. 
Thus, as was the case with the SPQE calculations, the maximum number of micro-iterations was set to 
50 to reduce the number of gradient evaluations.

All correlated approaches were based on RHF references with the one- and two-electron integrals 
obtained with Psi4 \cite{psi4}.

%

\clearpage
\renewcommand{\thefigure}{S\arabic{figure}}
\setcounter{figure}{0}
\setcounter{page}{1}
\renewcommand{\thepage}{S\arabic{page}}

\renewcommand{\thesection}{S\Roman{section}}
\setcounter{section}{0}

\onecolumngrid
\fontsize{12}{24}\selectfont
\begin{center}
	\textbf{\large Supplemental Material:\\
	CNOT-Efficient Circuits for Arbitrary Rank Many-Body Fermionic and Qubit Excitations
	}\\[.2cm]
	Ilias Magoulas$^{1,*}$ and Francesco A.\ Evangelista$^{1,*}$\\[.1cm]
	{\itshape ${}^1$Department of Chemistry and Cherry Emerson Center for Scientific Computation,\\ 
	Emory University, Atlanta, Georgia 30322, USA\\}
	${}^*$Corresponding authors; e-mails: ilias.magoulas@emory.edu (I.M.), 
	francesco.evangelista@emory.edu (F.A.E.).
\end{center}

\newpage
\noindent
This Supplemental Material document is organized as follows. In Sec.\ \ref{multi-qubit-controlled} 
we describe how to decompose the multi-qubit-controlled $R_y(2 \theta)$ gate, which is needed to 
define the 
CNOT-efficient quantum circuits of Figs.\ \ref{figure4} and \ref{figure5} in the main text, in 
terms of single- 
and two-qubit gates. In Sec.\ \ref{remove_1cnot}, we contemplate the use of circuit identities for 
the removal 
of a single CNOT gate from the quantum circuits of Figs.\ \ref{figure4} and \ref{figure5} in the 
main text. 
Section \ref{results} provides, in a graphical form, the results of our additional 
numerical simulations. In particular, in Sec.\ \ref{ucc_convergence} we compare the convergence to 
the exact, full configuration interaction (FCI), solution of 
the fermionic-excitation-based (FEB) and qubit-excitation-based (QEB) flavors of UCCSD, UCCSDT, and 
UCCSDTQ for the symmetric dissociation of the linear $\text{H}_6$/STO-6G system. In Sec.\ 
\ref{sm_sec_standard_vs_efficient} we compare the CNOT counts obtained with \spqe{1} and 
FEB-\spqe{1} 
and with q\spqe{1} and QEB-\spqe{1} for the symmetric dissociations of the linear $\text{BeH}_2$ 
and $\text{H}_6$ systems, the symmetric dissociation of the $\text{H}_6$ ring, and the insertion of 
Be into $\text{H}_2$, as described by the STO-6G basis. In the cases of the 
symmetric dissociation of the $\text{H}_6$ ring and the insertion of Be into 
$\text{H}_2$, we also include the pertinent results obtained with the tighter $\Omega = 
\SI{e-2}{\textit{E}_h}$ threshold. In Sec.\ \ref{sm_sec_feb_vs_qeb_spqe} we compare the efficiency 
of the FEB 
and QEB variants of \spqe{1} for the symmetric dissociations of the linear $\text{BeH}_2$ and 
$\text{H}_6$ systems, the symmetric dissociation of the $\text{H}_6$ ring, and the 
insertion of Be into $\text{H}_2$, as described by the STO-6G basis. In the cases of the 
symmetric dissociation of the $\text{H}_6$ ring and the insertion of Be into 
$\text{H}_2$, we also examine the performance of FEB- and QEB-\spqe{2}. In Sec.\ 
\ref{sec_feb_vs_qeb_adapt}, we compare the efficiency of the FEB and QEB variants of ADAPT-VQE for 
the symmetric dissociations of the linear $\text{BeH}_2$ and $\text{H}_6$ systems, the symmetric 
dissociation of the $\text{H}_6$ ring, and the insertion of Be into $\text{H}_2$, as 
described by the STO-6G basis. We examined two operator pools, namely, particle--hole singles and 
doubles and 
their generalized extension, and two convergence criteria, namely, $10^{-2}$ and 
\SI{e-3}{\textit{E}_h}. Finally, in Sec.\ \ref{sm_sec_feb-spqe_vs_qeb-adapt}, we compare the 
performance of 
FEB-\spqe{2} with that of QEB-\adaptsd{3} and its GSD variant for  the symmetric dissociation of 
the $\text{H}_6$ ring and the insertion of Be into $\text{H}_2$, as described by the 
STO-6G basis.

\newpage

\section{Decomposition of Multi-Qubit-Controlled $\boldsymbol{R_y (2 \theta)}$ Gate}
\label{multi-qubit-controlled}

\begin{figure*}[h!]
	\centering
	\includegraphics[width=\textwidth]{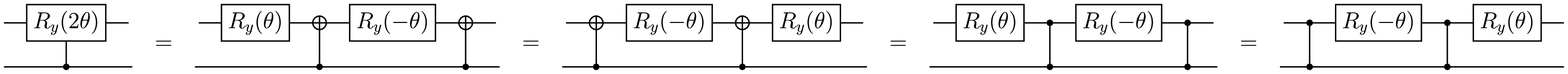}
	\caption{\label{fig_control_decomp}
		Decomposition of controlled $R_y (2 \theta)$ gate in terms of single-qubit 
		$R_y (\theta)$ and $R_y (-\theta)$ rotations and two-qubit CNOT/CZ gates.
	}
\end{figure*}

\begin{figure*}[h!]
	\centering
	\includegraphics[scale=0.5]{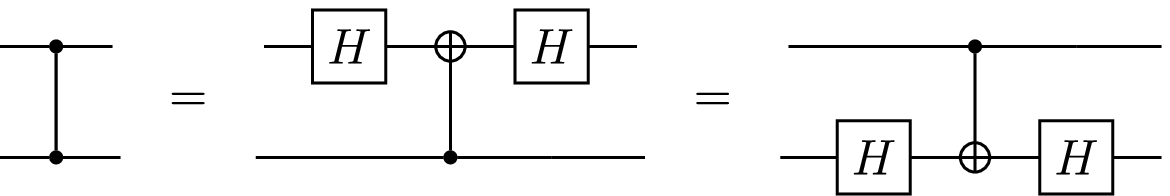}
	\caption{\label{fig_cz_to_cnot}
		Transformation of a CZ gate into a CNOT.
	}
\end{figure*}

\begin{figure*}[h!]
	\centering
	\includegraphics[width=\textwidth]{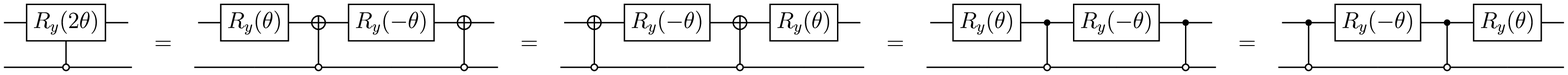}
	\caption{\label{fig_anti_decomp}
		Decomposition of anti-controlled $R_y (2 \theta)$ gate in terms of single-qubit 
		$R_y (\theta)$ and $R_y (-\theta)$ rotations and two-qubit anti-CNOT/anti-CZ gates.
	}
\end{figure*}

\begin{figure*}[h!]
	\centering
	\includegraphics[scale=0.5]{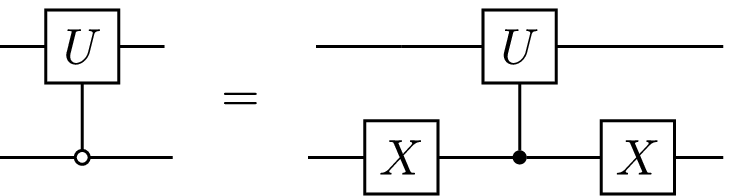}
	\caption{\label{fig_anti_to_control}
		Transformation of an anti-controlled gate into its controlled counterpart.
	}
\end{figure*}

Before we outline how to decompose a general multi-qubit-controlled $R_y (2 \theta)$ gate, we begin 
with the simpler case of an $R_y (2 \theta)$ rotation controlled on a single qubit. As shown 
in Refs.\ \cite{Barenco1995,Yordanov2019,Yordanov2020} of the main text
and Fig.\ \ref{fig_control_decomp}, there are multiple ways 
of expressing a controlled $R_y (2 \theta)$ rotation in terms of single- and two-qubit gates. What 
is common in all of these circuits is that the controlled  $R_y (2 \theta)$ gate is replaced by two 
half-way rotations of opposite directions, $R_y(\theta)$ and $R_y(-\theta)$, and two two-qubit 
gates, either CNOTs or CZs. There are two possible arrangements of the gates in the resulting 
circuits, namely, $R_y(\theta) \rightarrow \text{TQG} \rightarrow R_y(-\theta) \rightarrow 
\text{TQG}$ and $\text{TQG} \rightarrow R_y(-\theta) \rightarrow \text{TQG} \rightarrow 
R_y(\theta)$, with TQG being either CNOT or CZ. We will refer to these arrangements as 
``forward''/``backward'' since a two-qubit gate is placed after/before each $R_y$ rotation. 
Note that the CNOT gates appearing in Fig.\ \ref{fig_control_decomp} are controlled on the bottom 
qubit, as is the case with the controlled $R_y (2 \theta)$ gate. Nevertheless, the flexibility in 
transforming a CZ gate to a CNOT, as shown in Fig.\ \ref{fig_cz_to_cnot}, allows one to construct 
additional 
equivalent circuits where the CNOT gates are controlled on the top qubit. In the case of an 
anti-controlled $R_y (2 \theta)$ gate, there are two families of decompositions. The first one, 
shown in Fig.\ \ref{fig_anti_decomp}, relies on anti-CNOT/anti-CZ gates. In the second family of 
decompositions, using the circuit identity shown in Fig.\ \ref{fig_anti_to_control}, the 
anti-controlled $R_y (2 \theta)$ gate is transformed to its controlled analog and its decomposition 
proceeds as in Fig.\ \ref{fig_control_decomp}.

\FloatBarrier

\FloatBarrier

\FloatBarrier

\FloatBarrier

\begin{figure*}[h!]
	\centering
	\includegraphics[width=\textwidth]{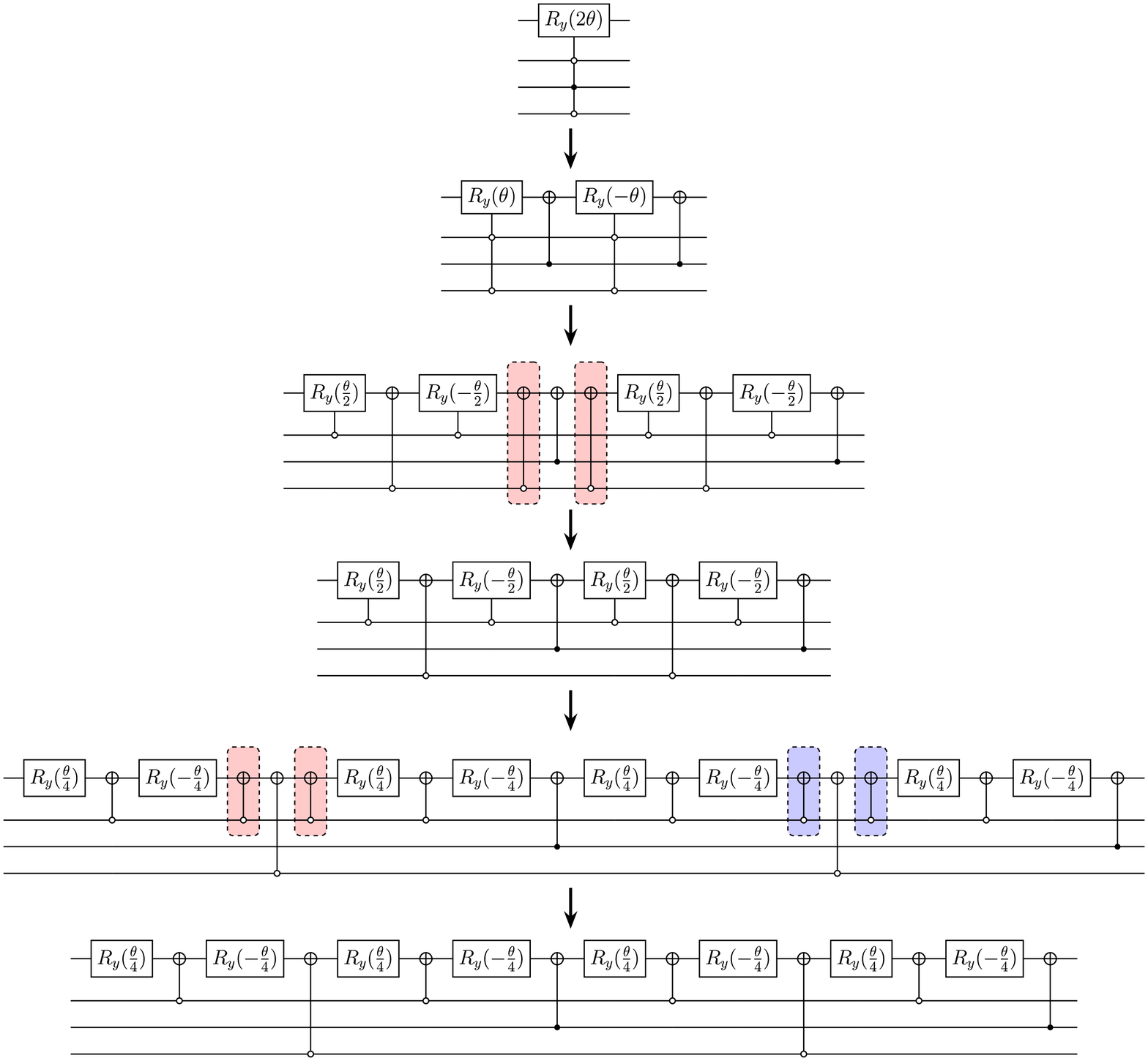}
	\caption{\label{fig_doubles_decomp}
		Decomposition of multi-qubit-controlled $R_y (2\theta)$ gate appearing 
		in the definition of FEB/QEB double excitations. CNOT gates shaded with
		the same color cancel out.
	}
\end{figure*}

The decomposition of a multi-qubit-controlled $R_y (2 \theta)$ gate is accomplished by applying the 
identities in Figs.\ \ref{fig_control_decomp} and \ref{fig_anti_decomp} one (anti-)control qubit at 
a time (see, for example, Ref.\ \cite{Yordanov2020} of the main text). For example, given an 
$n$-qubit-controlled $R_y (2 \theta)$ gate, we start by selecting one of the (anti-)control qubits. 
After 
the first decomposition, we end up with 2 ($n-1$)-qubit-controlled $R_y$ gates, with angles 
$\theta$ and $-\theta$, and 2 two-qubit gates. In the next step, we choose another (anti-)control 
qubit and decompose both of the ($n-1$)-qubit-controlled $R_y$ gates. After the second 
decomposition, the circuit contains 4 ($n-2$)-qubit-controlled $R_y$ rotations, 2 with angle 
$\tfrac{\theta}{2}$ and 2 with angle $-\tfrac{\theta}{2}$, and 6 two-qubit gates. After repeating 
this procedure for all remaining (anti-)control qubits, the 
final quantum circuit will contain $2^{n}$ single-qubit $R_y$ rotations, half with angle 
$\tfrac{\theta}{2^{n-1}}$ and half with angle $-\tfrac{\theta}{2^{n-1}}$, and $2^{n+1} - 2$ 
two-qubit gates. In the optimum implementation, one decomposes the multi-qubit-controlled $R_y$ 
rotations in a way that maximizes the number of adjacent equivalent two-qubit gates. This can be 
accomplished by alternating the ``forward'' and ``backward'' arrangements of decomposed 
multi-qubit-controlled $R_y$ rotations. By doing so, one cancels adjacent equivalent two-qubit 
gates, reducing their total number to $2^{n-1}$. In Fig.\ \ref{fig_doubles_decomp}, we illustrate 
the aforementioned procedure in the case of the multi-qubit-controlled $R_y (2\theta)$ gate 
appearing in the definition of FEB/QEB double excitations.

\FloatBarrier

\clearpage

\section{On the Use of Circuit Identities for the Removal of a Single CNOT Gate}
\label{remove_1cnot}

\begin{figure*}[h!]
	\centering
	\includegraphics[scale=0.5]{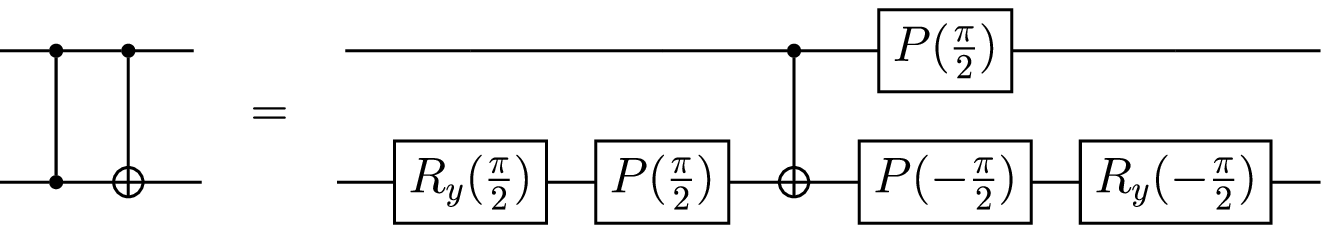}
	\caption{\label{fig_cz_times_cnot}
		Circuit identity that replaces the product of a CZ and a CNOT gate by a circuit containing 
		a single CNOT and single-qubit gates. Note that $P(\theta) \equiv
		\big(\begin{smallmatrix}
			1 & 0 \\
			0 & e^{i\theta}
		\end{smallmatrix} \big)$
		is the phase shift gate.
	}
\end{figure*}

\begin{figure*}[h!]
	\centering
	\includegraphics[width=\textwidth]{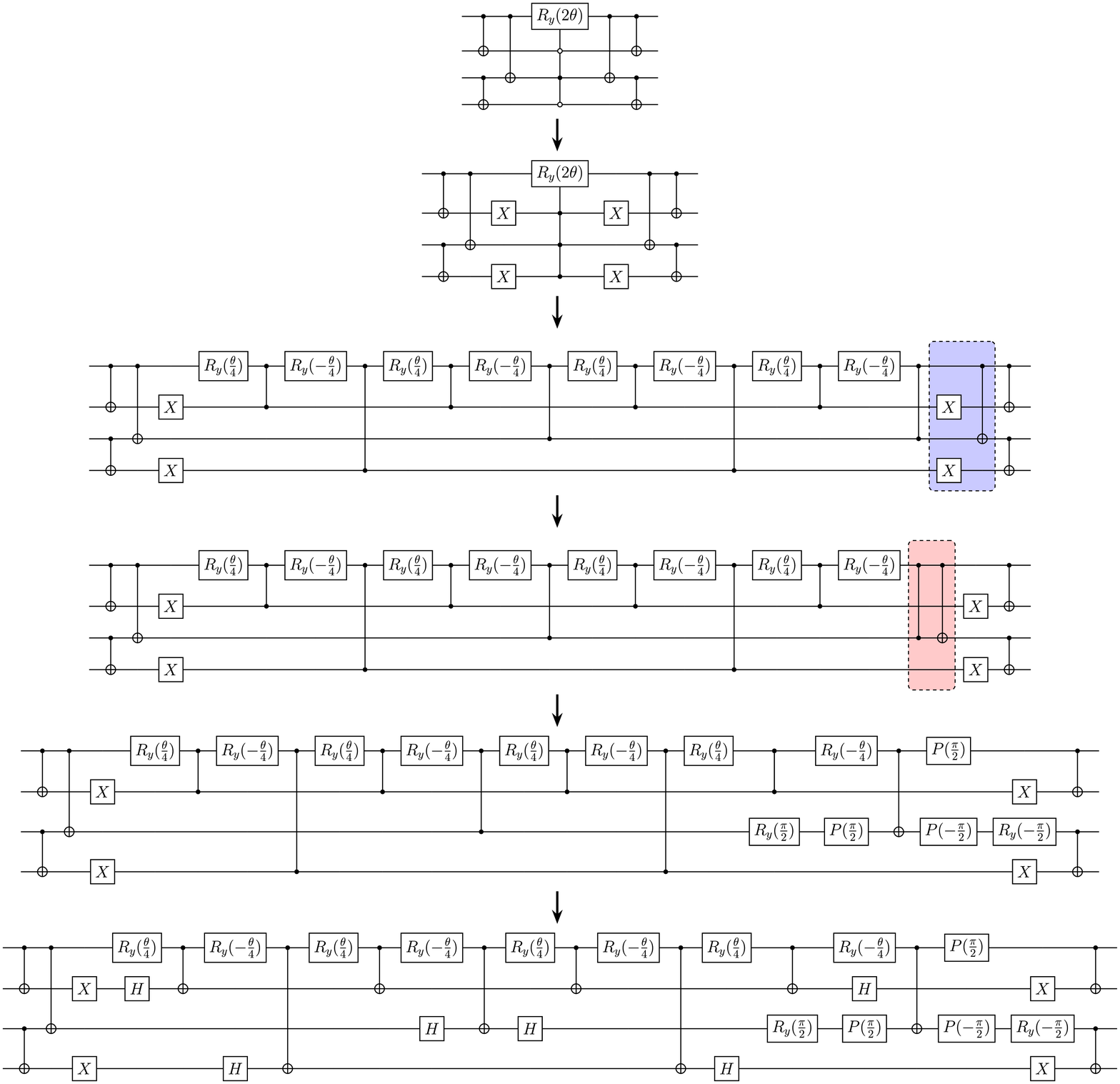}
	\caption{\label{fig_cnot_reduction_doubles}
		Illustration of the process that reduces the number of CNOT gates by 1, using the QEB 
		circuit that performs double qubit excitations as an example. The gates shaded in blue
		color commute. The circuit identity of Fig.\ \ref{fig_cz_times_cnot} is applied to the
		gates shaded in red color. Note that $P(\theta) \equiv
		\big(\begin{smallmatrix}
			1 & 0 \\
			0 & e^{i\theta}
		\end{smallmatrix} \big)$
		is the phase shift gate.
	}
\end{figure*}

In Ref.\ \cite{Yordanov2020} of the main text, Yordanov \foreign{et al.} demonstrated how to reduce 
the CNOT 
counts of the FEB and QEB quantum circuits by a single CNOT gate. The circuit identity behind this 
reduction is shown in Fig.\ \ref{fig_cz_times_cnot} and it involves the replacement of the product 
of a CZ gate times a CNOT by a single CNOT gate and five single-qubit ones. To take advantage of 
this circuit identity, the following procedure is typically employed. In the initial step, one uses 
the circuit identity of Fig.\ \ref{fig_anti_to_control} to replace all anti-control qubits of the 
multi-qubit-controlled $R_y (2\theta)$ gate by control ones. Subsequently, one decomposes the 
multi-qubit-controlled $R_y (2\theta)$ gate into a series of alternating $R_y$ rotations and CZ 
gates, similarly to the procedure outlined in Fig.\ \ref{fig_doubles_decomp}; note that the use of 
CZ 
gates rather than CNOTs is crucial for eliminating a CNOT gate. In the next step, one uses the 
circuit identity of Fig.\ \ref{fig_cz_times_cnot} to replace the right-most CZ gate and one of the 
CNOTs immediately after the multi-qubit-controlled $R_y(2\theta)$ gate by a single CNOT and five 
single-qubit gates. Depending on which two-qubit gates are directly available in a given hardware 
implementation, a final step might be involved in which the remaining CZ gates are transformed into 
CNOTs using the circuit identity of Fig.\ \ref{fig_cz_to_cnot}. An illustration of the above 
procedure can be found in Fig.\ \ref{fig_cnot_reduction_doubles}, using as an example the QEB 
quantum circuit performing a double qubit excitation.

As already mentioned, depending on the native gate set available on the hardware, one might need to 
replace all 
CZ gates by 
CNOTs. However, by doing so one introduces a number of Hadamard gates that scales linearly with the 
excitation rank $n$. To be precise, the number of required Hadamard gates is $4n - 2$. In total, 
the number of single-qubit gates increases by $4(n+1)$ when applying the identity responsible for 
eliminating a single CNOT. Therefore, as the excitation rank increases, the gate errors associated 
with the aforementioned single-qubit gates are anticipated to outweigh the benefits of removing a 
single CNOT.

\FloatBarrier

\FloatBarrier

\clearpage

\section{Results of Additional Numerical Simulations}
\label{results}

\subsection{Convergence to FCI of QEB-UCCSD, QEB-UCCSDT, and QEB-UCCSDTQ}
\label{ucc_convergence}

\begin{figure*}[h!]
	\centering
	\includegraphics[width=\textwidth]{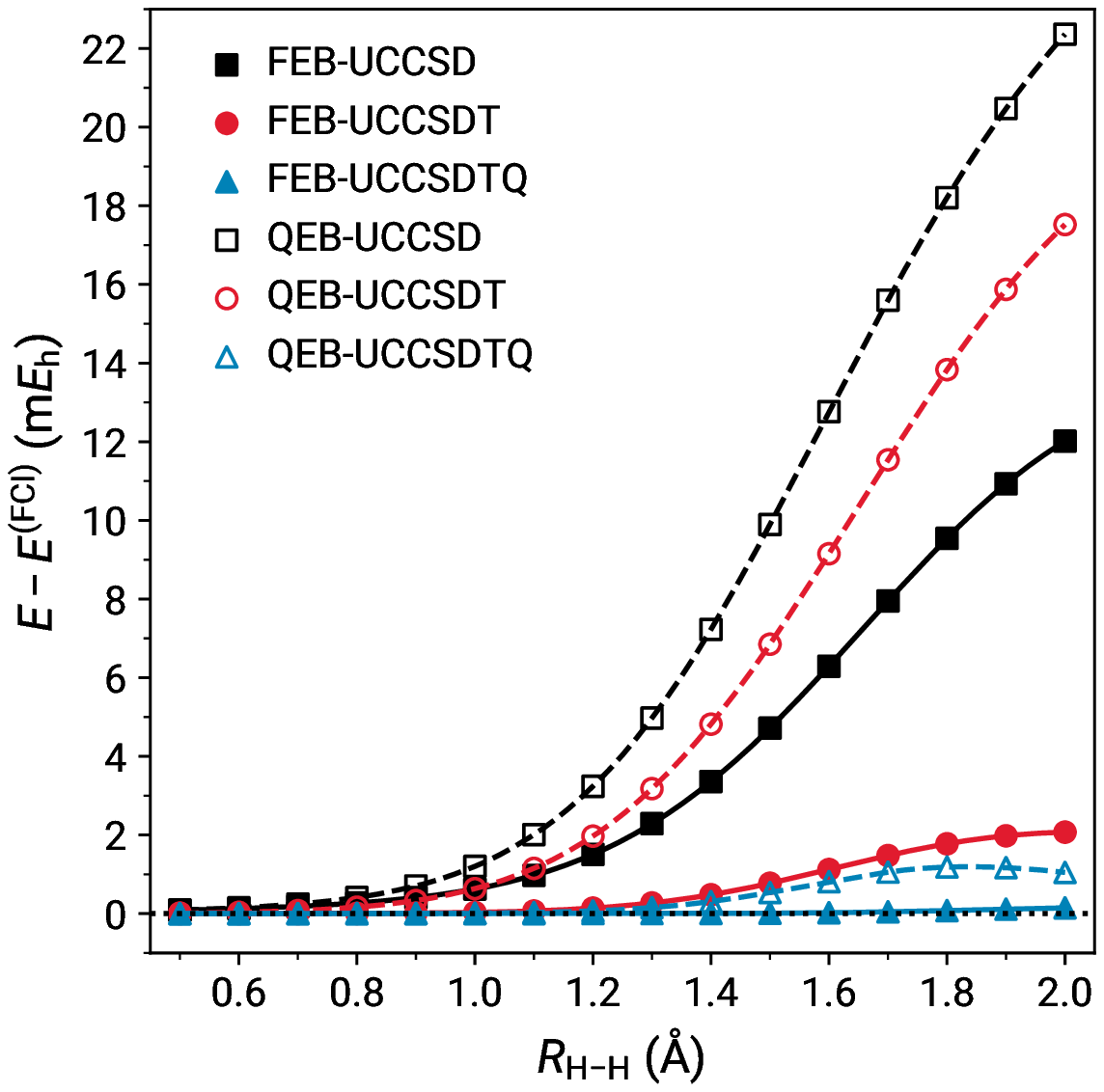}
	\caption{
		Errors relative to FCI characterizing the FEB-UCCSD, FEB-UCCSDT, FEB-UCCSDTQ, QEB-UCCSD, 
		QEB-UCCSDT, and QEB-UCCSDTQ simulations of the symmetric dissociation of the $\text{H}_6$ 
		linear chain as described by the STO-6G basis.
	}
\end{figure*}
\FloatBarrier

\clearpage

\subsection{Standard \foreign{vs} CNOT-Efficient SPQE}
\label{sm_sec_standard_vs_efficient}

\begin{figure*}[h!]
	\centering
	\includegraphics[width=0.92\textwidth]{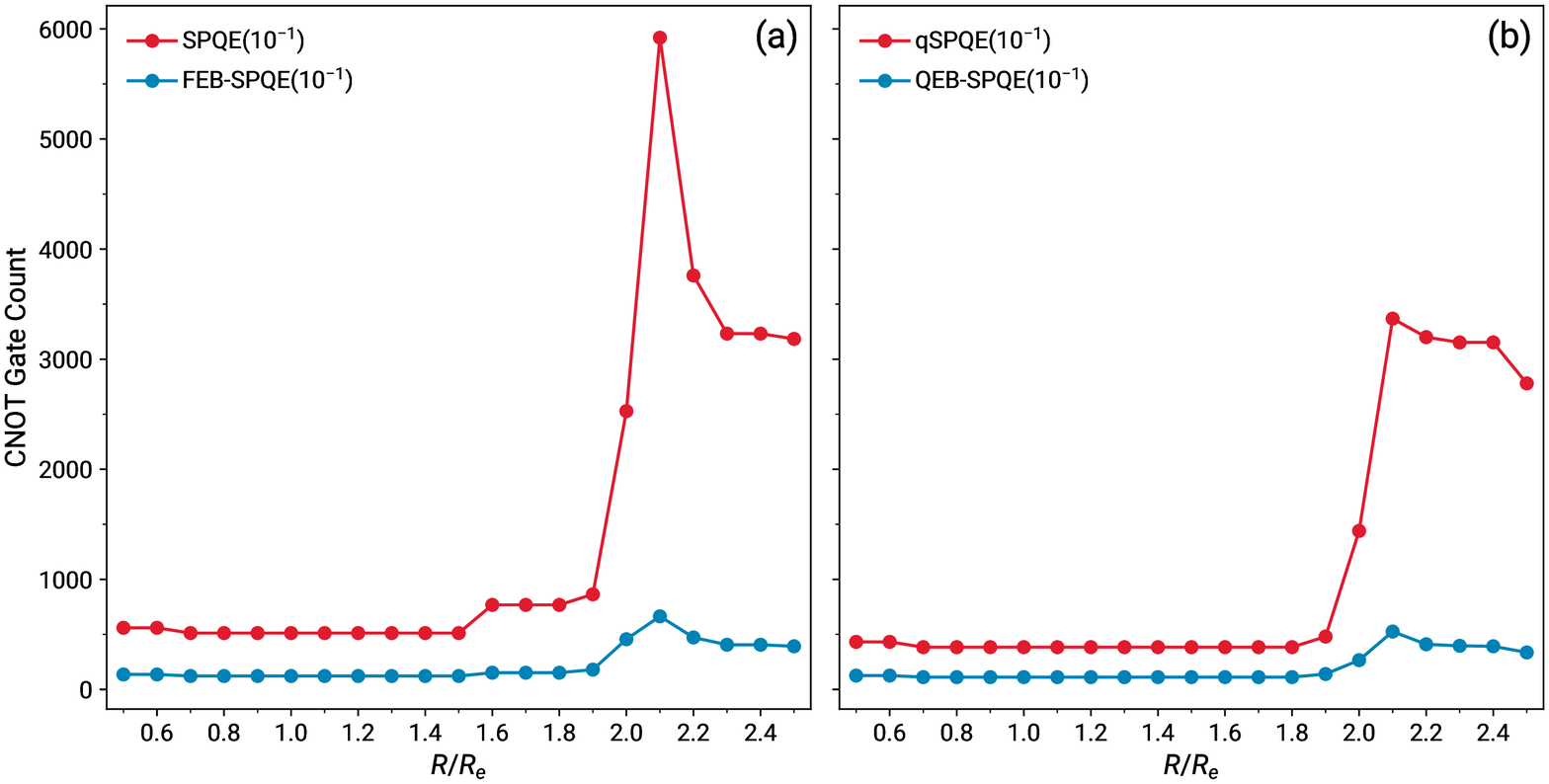}
	\caption{
		Total CNOT gate counts characterizing the (a) fermionic SPQE and FEB-SPQE and (b) qubit 
		qSPQE and QEB-SPQE ansatz unitaries for the symmetric dissociation of the linear
		$\text{BeH}_2$/STO-6G system. All computations employ a selection threshold of $\Omega = 
		\SI{e-1}{\textit{E}_h}$.
	}
\end{figure*}

\begin{figure*}[h!]
	\centering
	\includegraphics[width=0.92\textwidth]{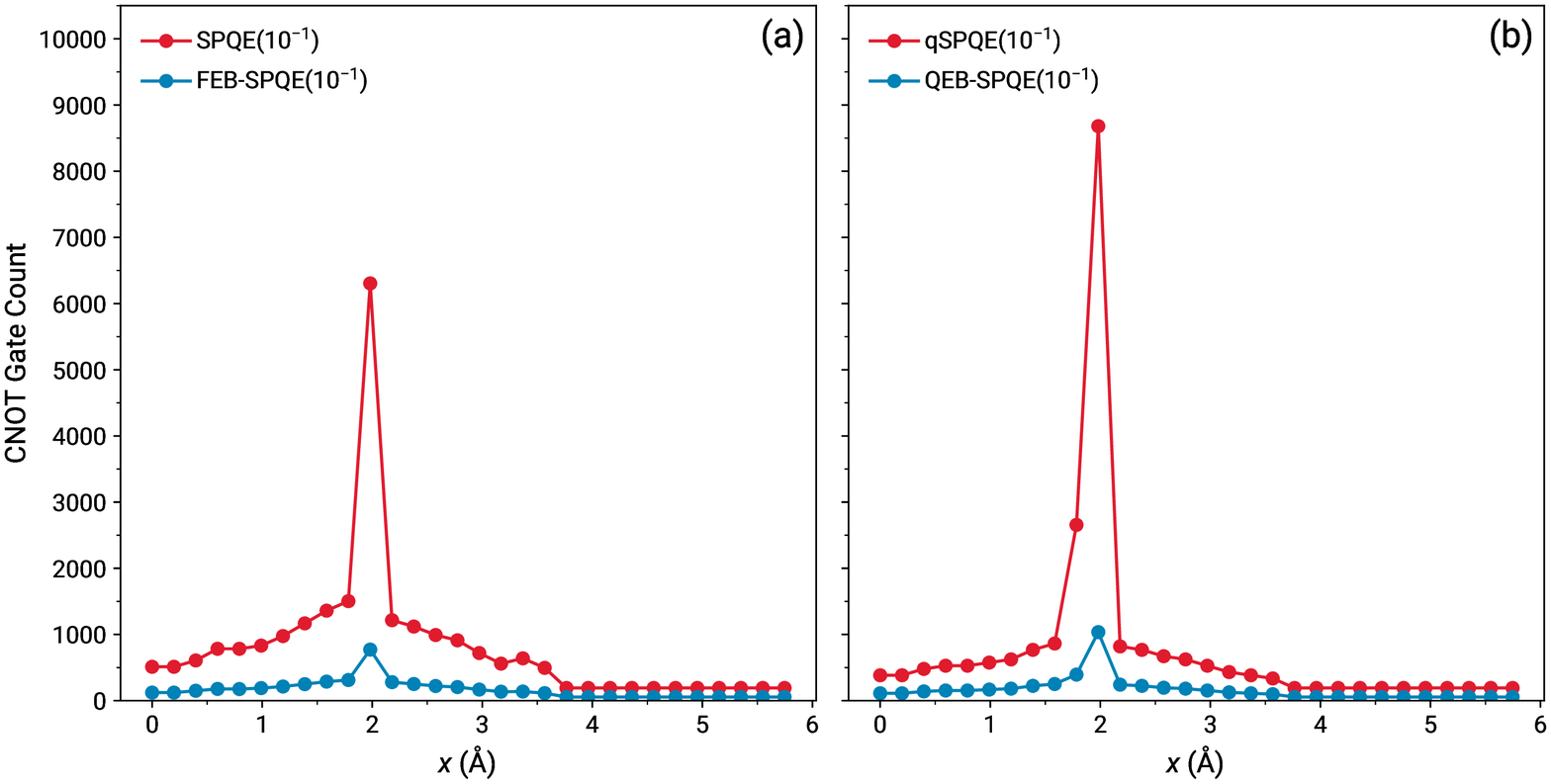}
	\caption{
		Total CNOT gate counts characterizing the (a) fermionic SPQE and FEB-SPQE and (b) qubit 
		qSPQE and QEB-SPQE ansatz unitaries for the insertion of Be into $\text{H}_2$. All 
		computations
		employ a selection threshold of $\Omega = \SI{e-1}{\textit{E}_h}$.
	}
\end{figure*}

\begin{figure*}[h!]
	\centering
	\includegraphics[width=\textwidth]{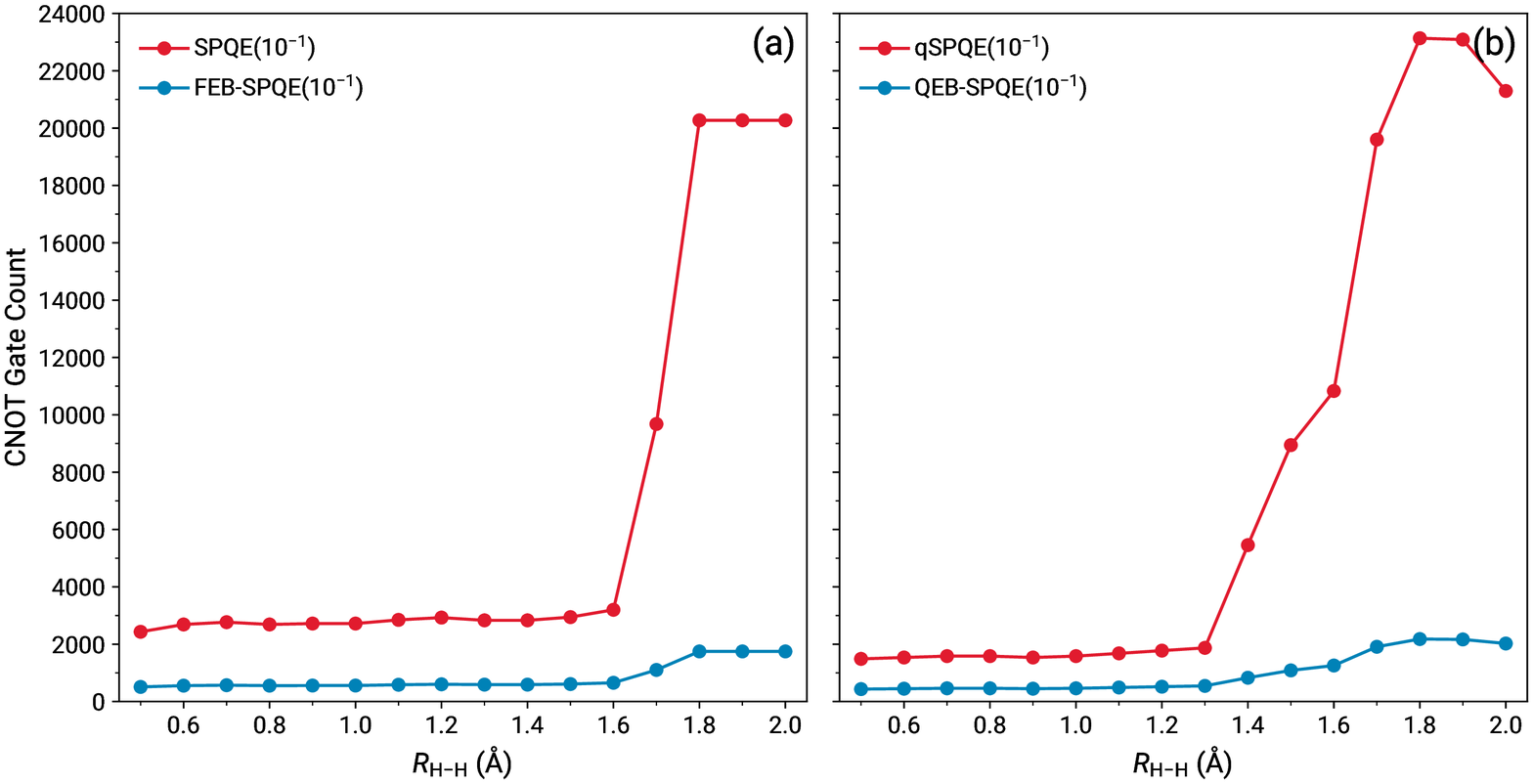}
	\caption{
		Total CNOT gate counts characterizing the (a) fermionic SPQE and FEB-SPQE and (b) qubit 
		qSPQE and QEB-SPQE ansatz unitaries for the symmetric dissociation of the linear 
		$\text{H}_6$/STO-6G 
		system. All computations employ a selection threshold of $\Omega = \SI{e-1}{\textit{E}_h}$.
	}
\end{figure*}

\begin{figure*}[h!]
	\centering
	\includegraphics[width=\textwidth]{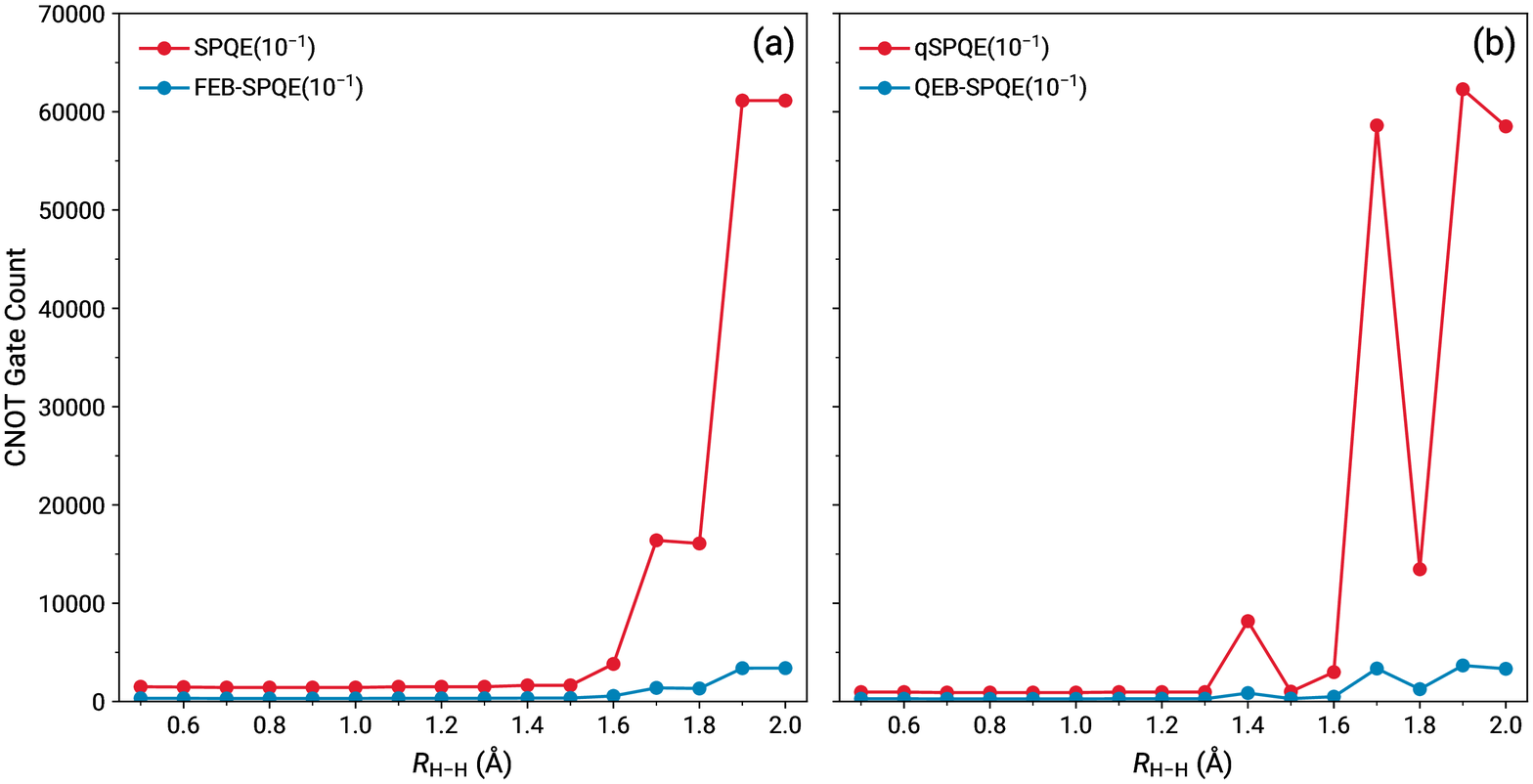}
	\caption{
		Total CNOT gate counts characterizing the (a) fermionic SPQE and FEB-SPQE and (b) qubit 
		qSPQE and QEB-SPQE ansatz unitaries for the symmetric dissociation of the 
		$\text{H}_6$/STO-6G ring.
		All computations employ a selection threshold of $\Omega = \SI{e-1}{\textit{E}_h}$.
	}
\end{figure*}

\begin{figure*}[h!]
	\centering
	\includegraphics[width=\textwidth]{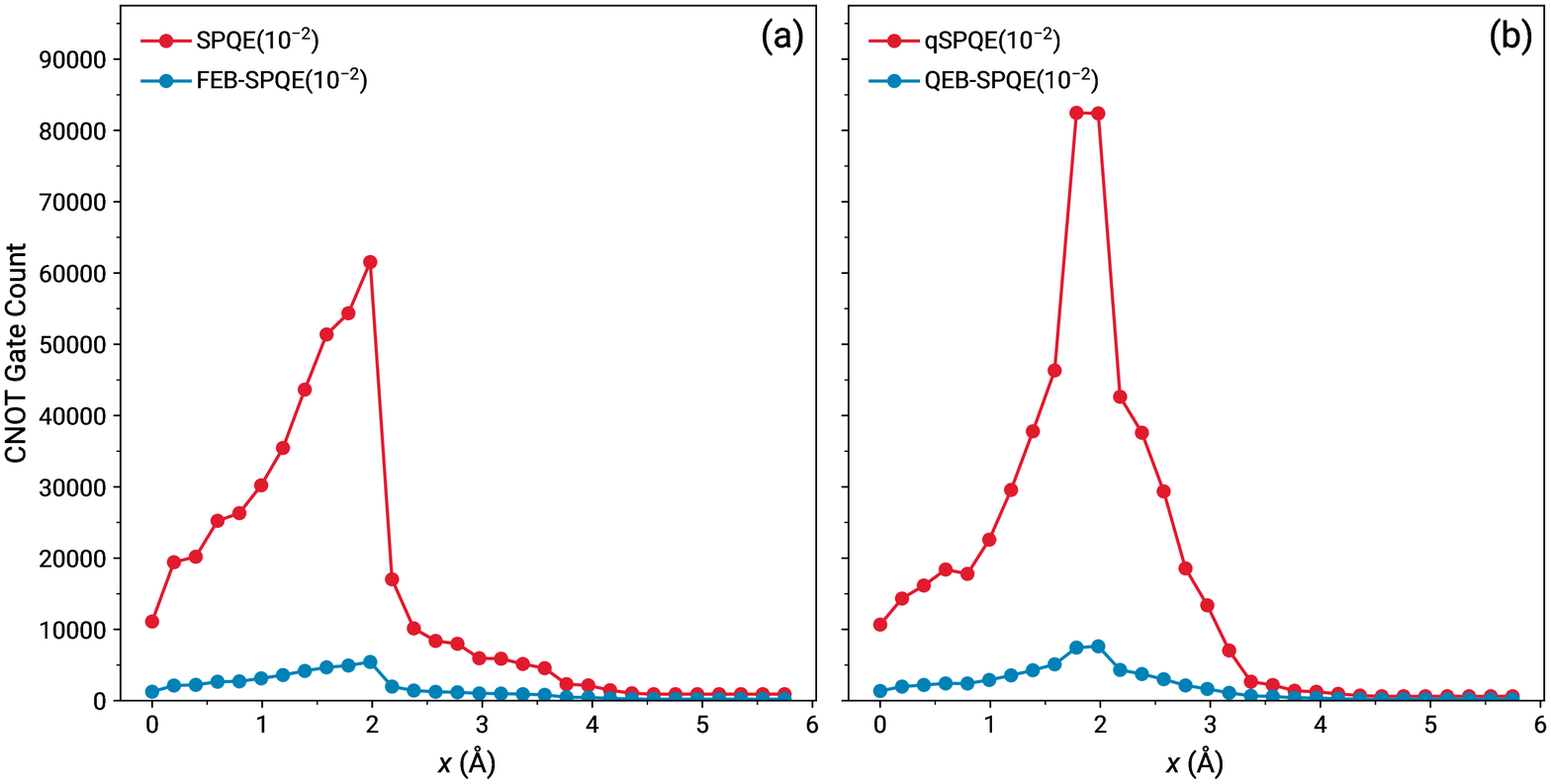}
	\caption{
		Total CNOT gate counts characterizing the (a) fermionic SPQE and FEB-SPQE and (b) qubit 
		qSPQE and QEB-SPQE ansatz unitaries for the insertion of Be into $\text{H}_2$. All 
		computations employ a 
		selection threshold of $\Omega = \SI{e-2}{\textit{E}_h}$.
	}
\end{figure*}

\begin{figure*}[h!]
	\centering
	\includegraphics[width=\textwidth]{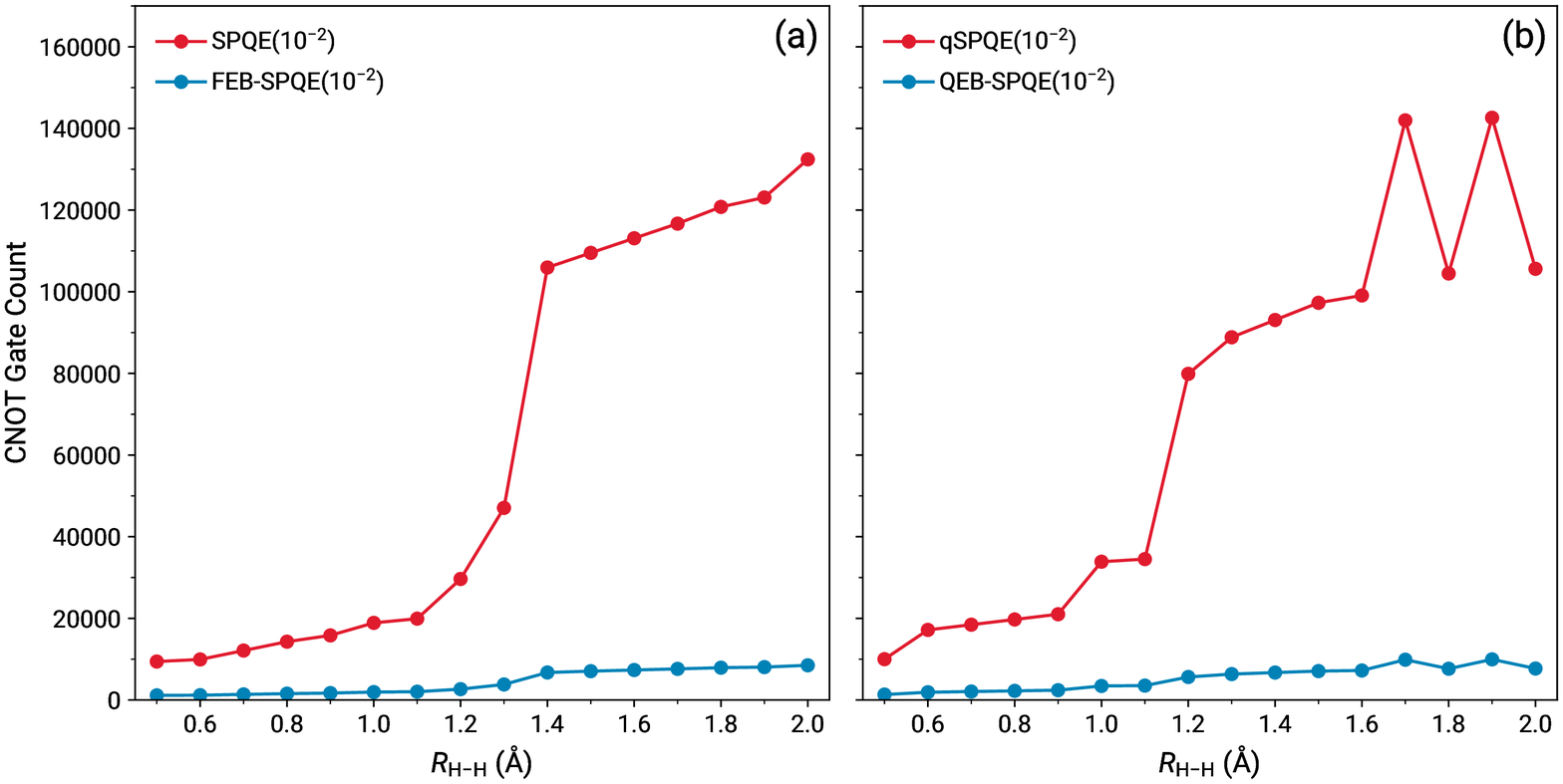}
	\caption{
		Total CNOT gate counts characterizing the (a) fermionic SPQE and FEB-SPQE and (b) qubit 
		qSPQE and QEB-SPQE ansatz unitaries for the symmetric dissociation of the 
		$\text{H}_6$/STO-6G ring. All 
		computations employ a selection threshold of $\Omega = \SI{e-2}{\textit{E}_h}$.
	}
\end{figure*}
\FloatBarrier

\subsection{FEB-SPQE \foreign{vs} QEB-SPQE}
\label{sm_sec_feb_vs_qeb_spqe}

\begin{figure*}[h!]
	\centering
	\includegraphics[height=0.78\textheight]{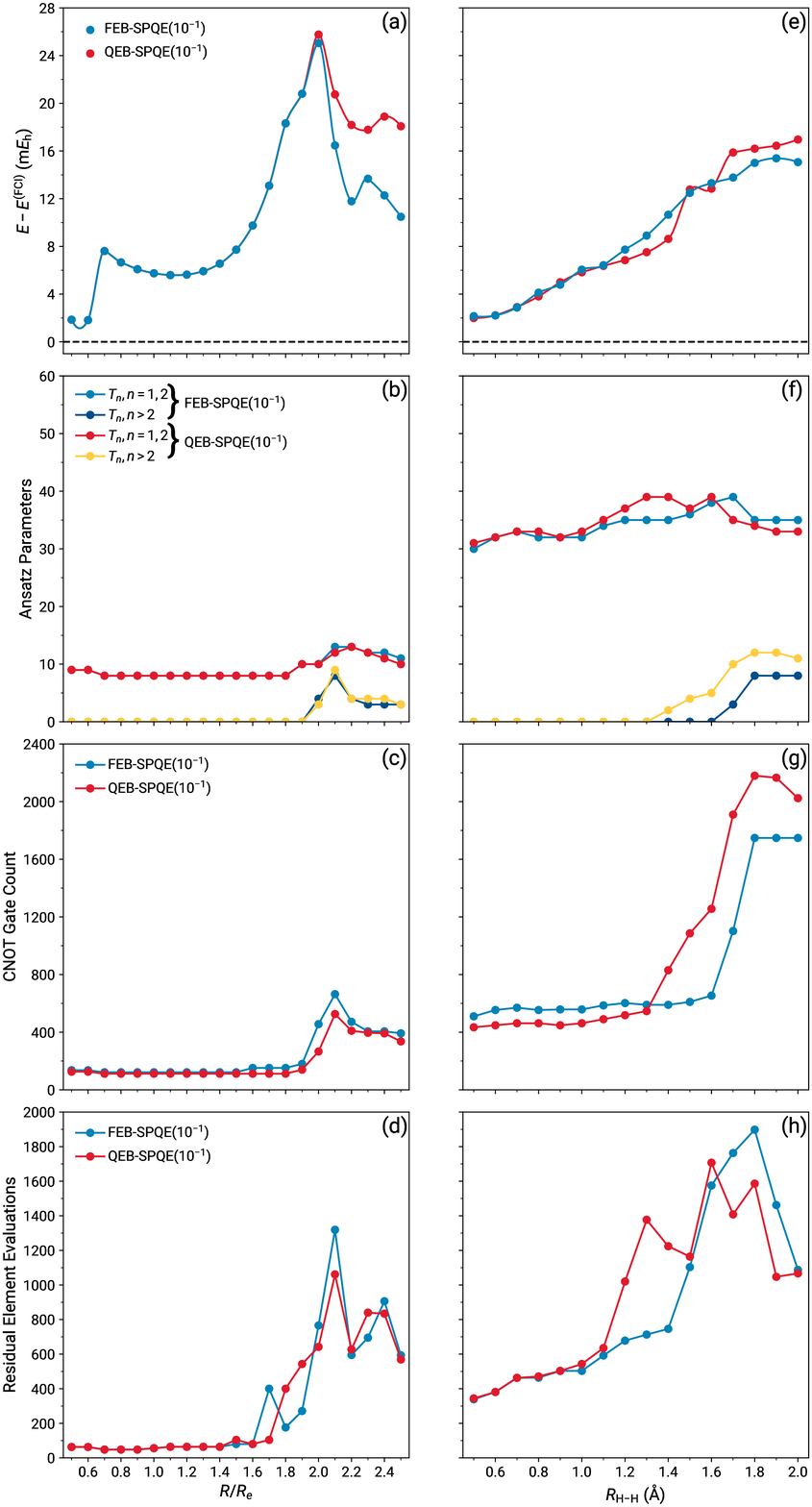}
	\caption{
		Errors relative to FCI [(a) and (e)], ansatz parameters [(b) and (f)], CNOT gate counts 
		[(c) and (g)], and residual element evaluations [(d) and (h)] characterizing the FEB- and 
		QEB-\spqe{1} simulations of the symmetric dissociations of the linear
		$\text{BeH}_2$ (left column) and $\text{H}_6$ (right column) systems as described by the 
		STO-6G basis.
	}
\end{figure*}

\begin{figure*}[h!]
	\centering
	\includegraphics[height=0.85\textheight]{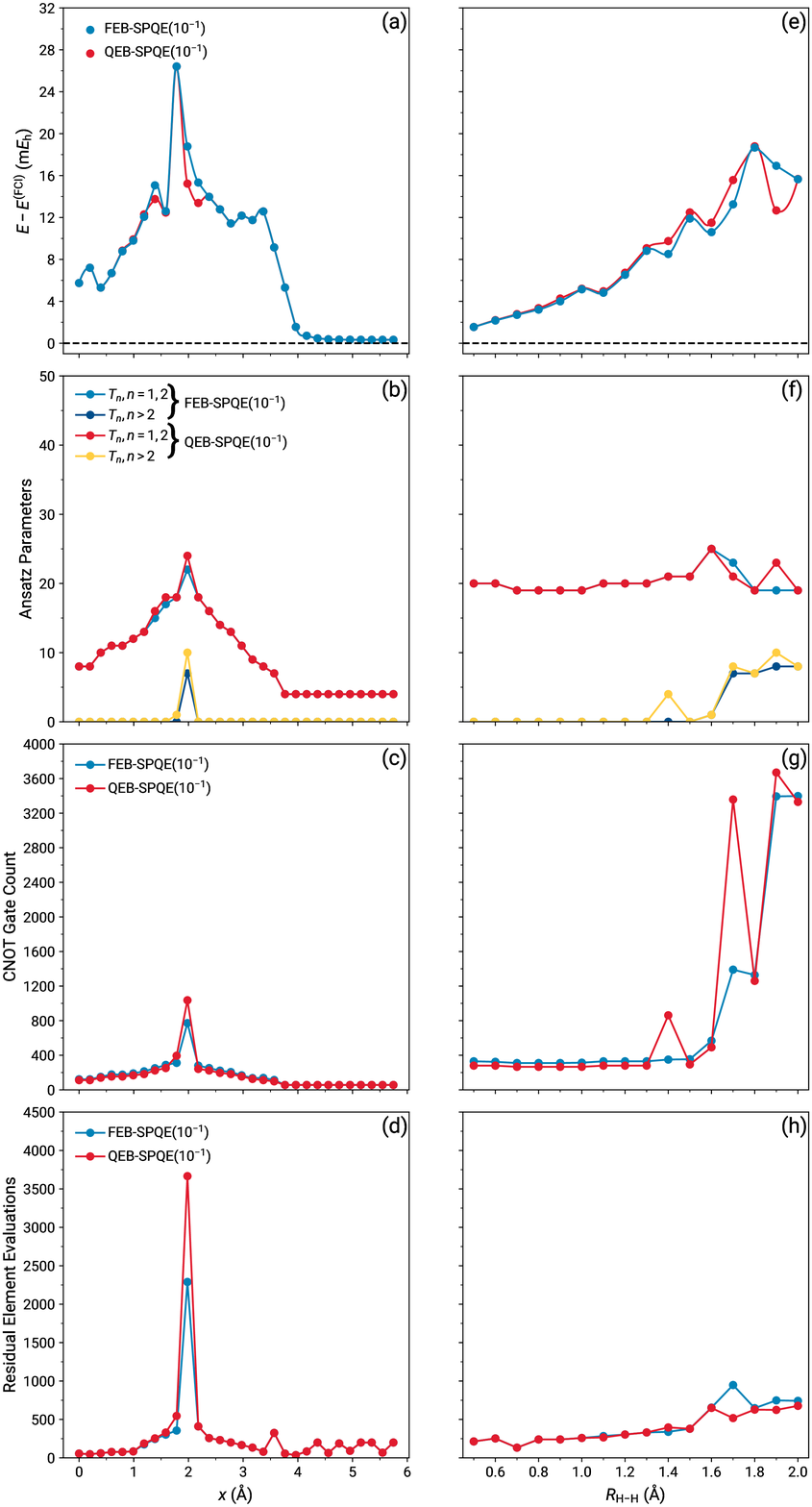}
	\caption{
		Errors relative to FCI [(a) and (e)], ansatz parameters [(b) and (f)], CNOT gate counts 
		[(c) and (g)], and residual element evaluations [(d) and (h)] characterizing the FEB- and 
		QEB-\spqe{1} simulations of the insertion of Be to $\text{H}_2$ (left 
		column) and the symmetric dissociation of the $\text{H}_6$ ring (right column) as 
		described by the STO-6G basis.
	}
\end{figure*}

\begin{figure*}[h!]
	\centering
	\includegraphics[height=0.85\textheight]{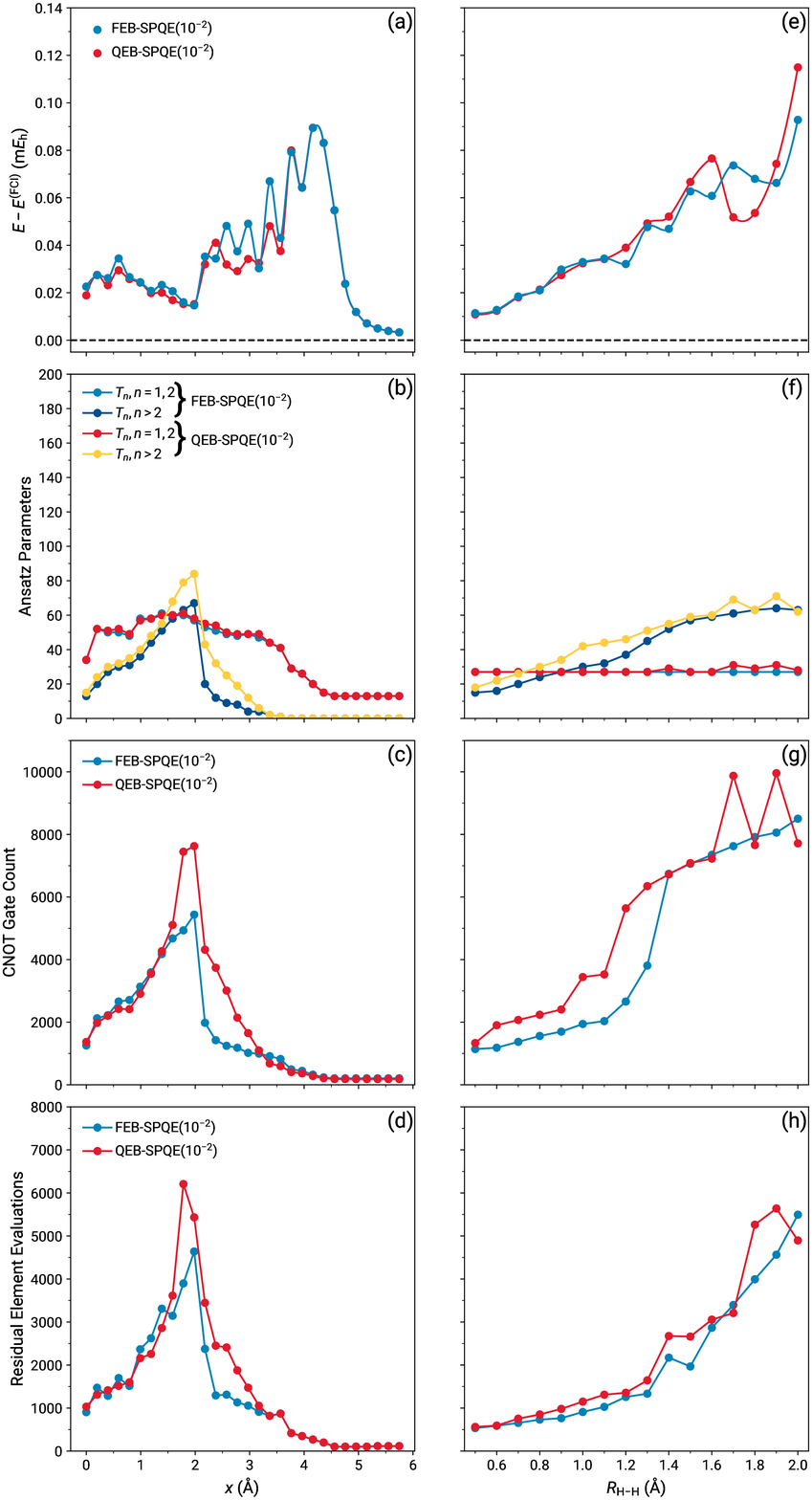}
	\caption{
		Errors relative to FCI [(a) and (e)], ansatz parameters [(b) and (f)], CNOT gate counts 
		[(c) and (g)], and residual element evaluations [(d) and (h)] characterizing the FEB- and 
		QEB-\spqe{2} simulations of the insertion of Be to $\text{H}_2$ (left 
		column) and the symmetric dissociation of the $\text{H}_6$ ring (right column) as 
		described by the STO-6G basis.
	}
\end{figure*}
\FloatBarrier

\subsection{FEB-ADAPT-VQE \foreign{vs} QEB-ADAPT-VQE}
\label{sec_feb_vs_qeb_adapt}

\begin{figure*}[h!]
	\centering
	\includegraphics[height=0.78\textheight]{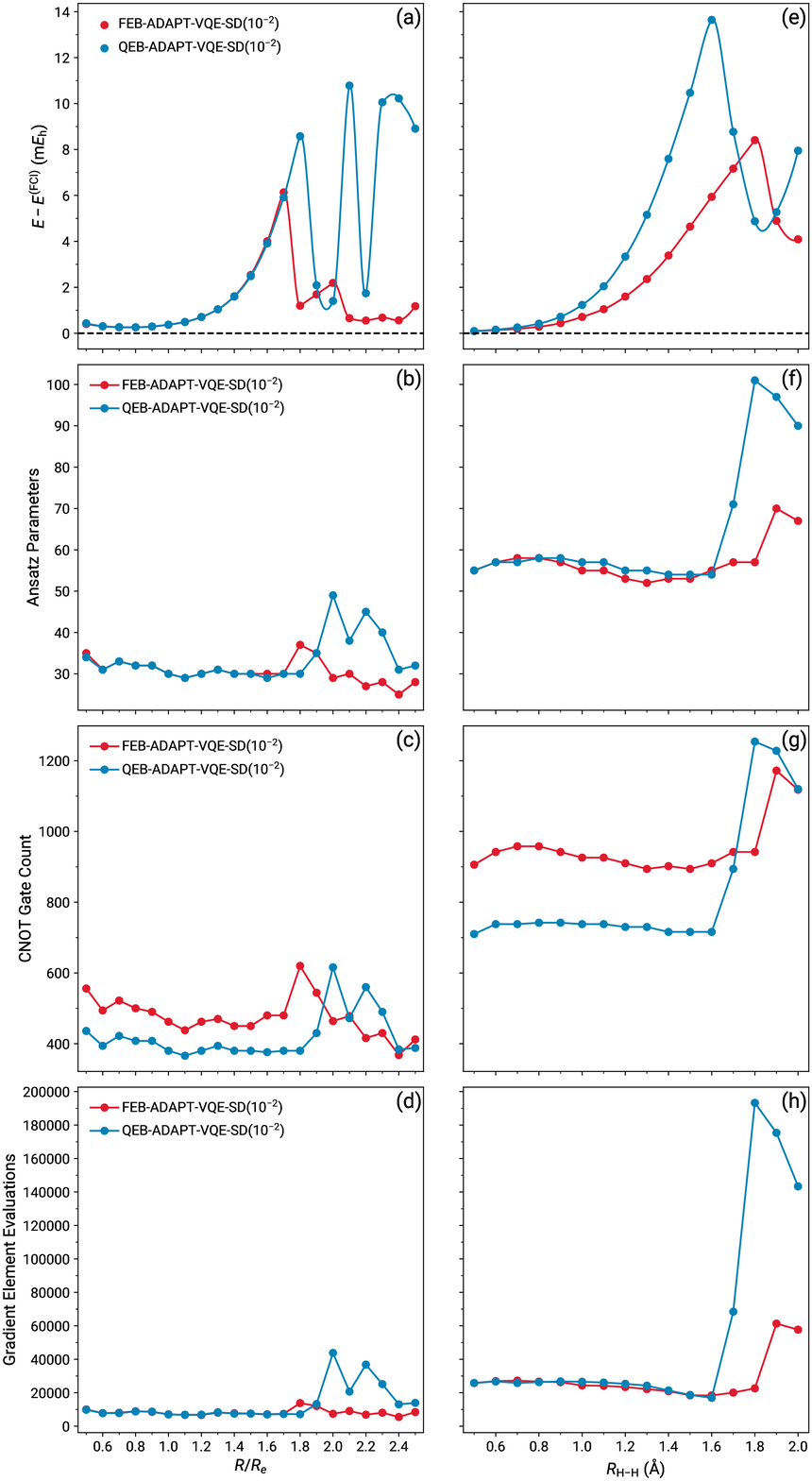}
	\caption{
		Errors relative to FCI [(a) and (e)], ansatz parameters [(b) and (f)], CNOT gate counts 
		[(c) and (g)], and gradient element evaluations [(d) and (h)] characterizing 
		the FEB- and QEB-\adaptsd{2} simulations of the symmetric 
		dissociations of the linear $\text{BeH}_2$ (left column) and $\text{H}_6$ (right column) 
		systems as described by the STO-6G basis.
	}
\end{figure*}

\begin{figure*}[h!]
	\centering
	\includegraphics[height=0.85\textheight]{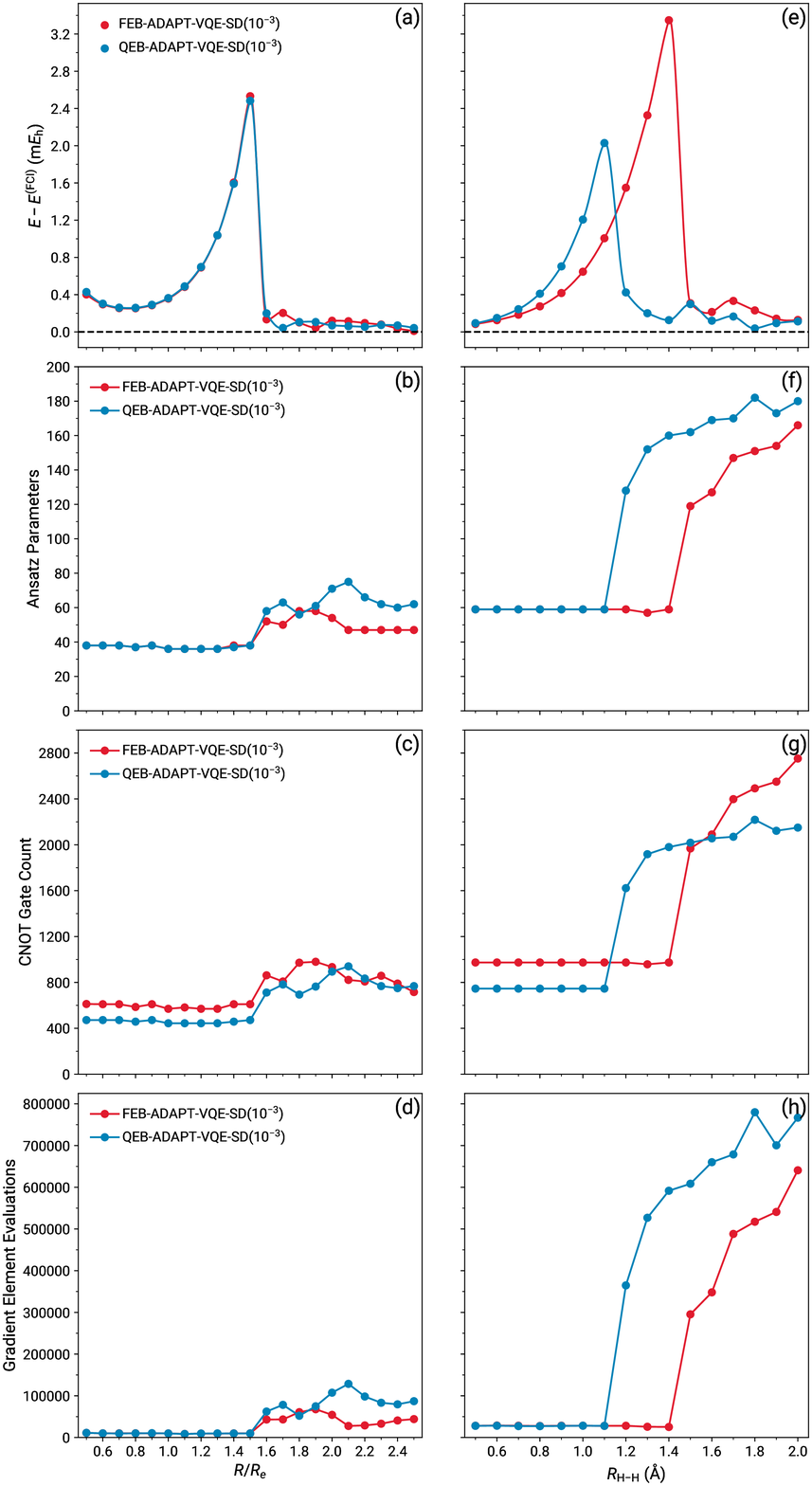}
	\caption{
		Errors relative to FCI [(a) and (e)], ansatz parameters [(b) and (f)], CNOT gate counts 
		[(c) and (g)], and gradient element evaluations [(d) and (h)] characterizing 
		the FEB- and QEB-\adaptsd{3} simulations of the symmetric 
		dissociations of the linear $\text{BeH}_2$ (left column) and $\text{H}_6$ (right column) 
		systems as described by the STO-6G basis.
	}
\end{figure*}

\begin{figure*}[h!]
	\centering
	\includegraphics[height=0.85\textheight]{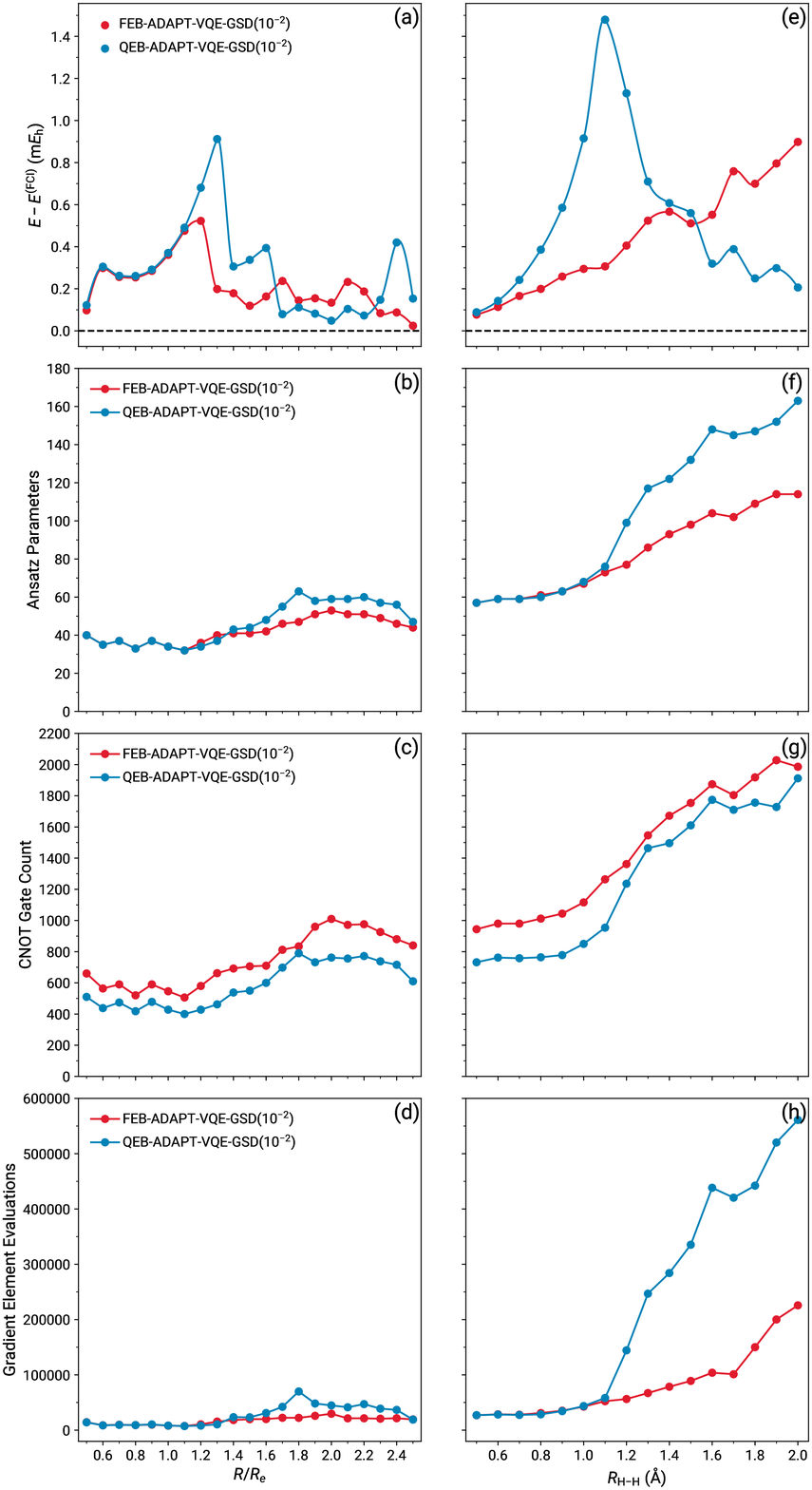}
	\caption{
		Errors relative to FCI [(a) and (e)], ansatz parameters [(b) and (f)], CNOT gate counts 
		[(c) and (g)], and gradient element evaluations [(d) and (h)] characterizing 
		the FEB- and QEB-\adaptgsd{2} simulations of the symmetric 
		dissociations of the linear $\text{BeH}_2$ (left column) and $\text{H}_6$ (right column) 
		systems as described by the STO-6G basis.
	}
\end{figure*}

\begin{figure*}[h!]
	\centering
	\includegraphics[height=0.85\textheight]{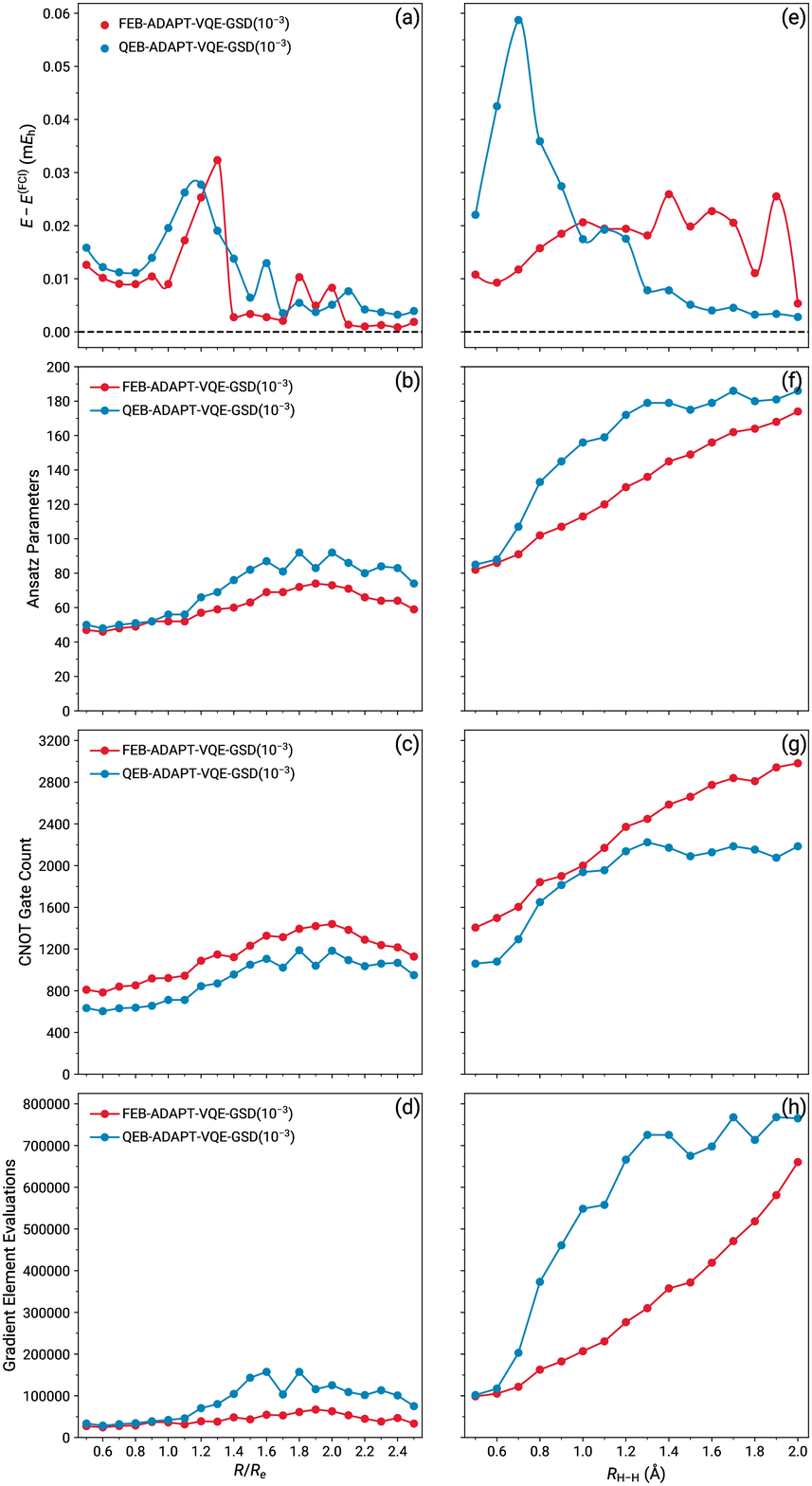}
	\caption{
		Errors relative to FCI [(a) and (e)], ansatz parameters [(b) and (f)], CNOT gate counts 
		[(c) and (g)], and gradient element evaluations [(d) and (h)] characterizing 
		the FEB- and QEB-\adaptgsd{3} simulations of the symmetric 
		dissociations of the linear $\text{BeH}_2$ (left column) and $\text{H}_6$ (right column) 
		systems as described 
		by the STO-6G basis.
	}
\end{figure*}

\begin{figure*}[h!]
	\centering
	\includegraphics[height=0.85\textheight]{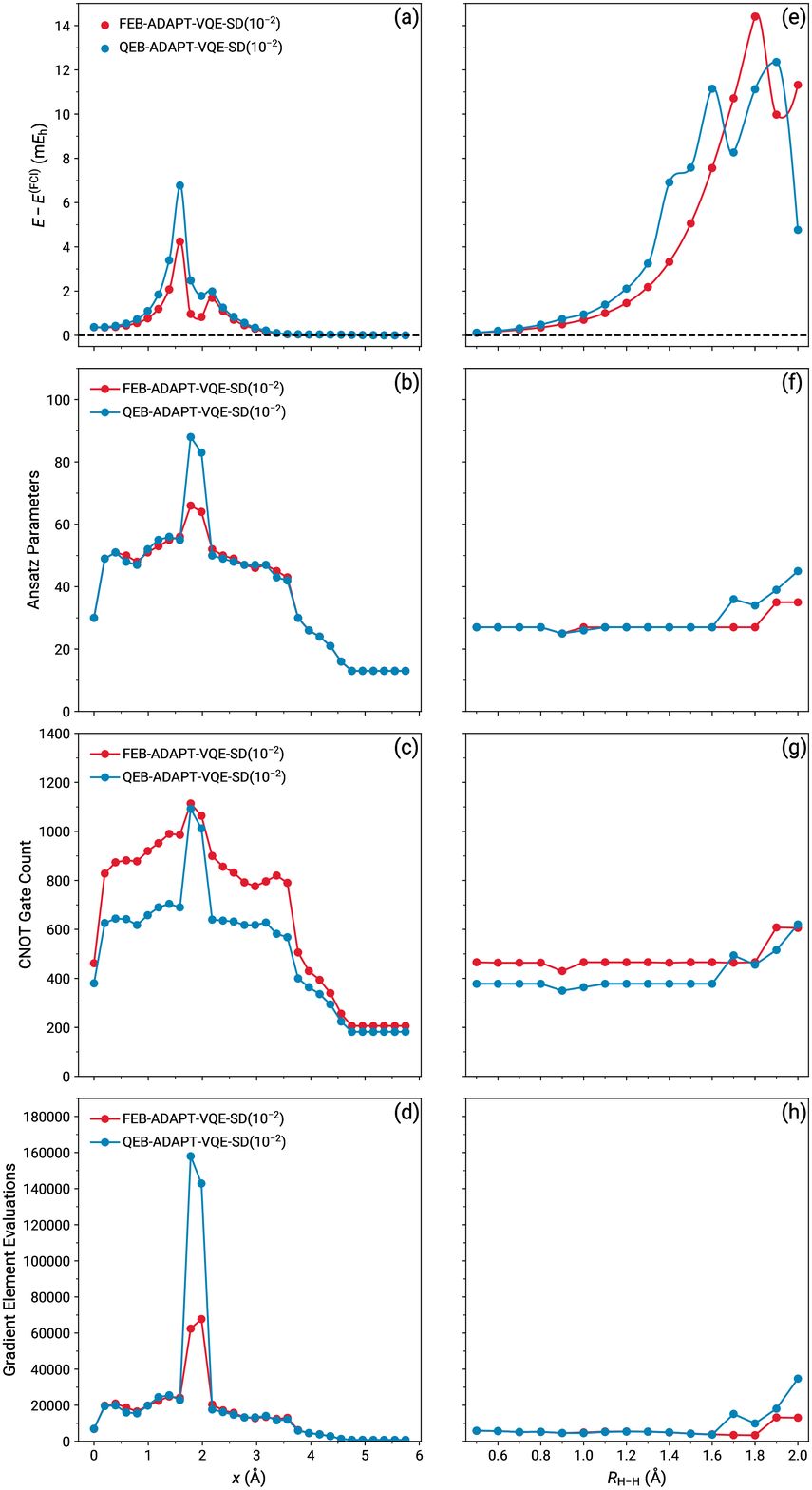}
	\caption{
		Errors relative to FCI [(a) and (e)], ansatz parameters [(b) and (f)], CNOT gate counts 
		[(c) and (g)], and gradient element evaluations [(d) and (h)] characterizing the FEB- and 
		QEB-\adaptsd{2} simulations of the insertion of Be to $\text{H}_2$ (left 
		column) and the symmetric dissociation of the $\text{H}_6$ ring (right column) as 
		described by the STO-6G basis.
	}
\end{figure*}

\begin{figure*}[h!]
	\centering
	\includegraphics[height=0.85\textheight]{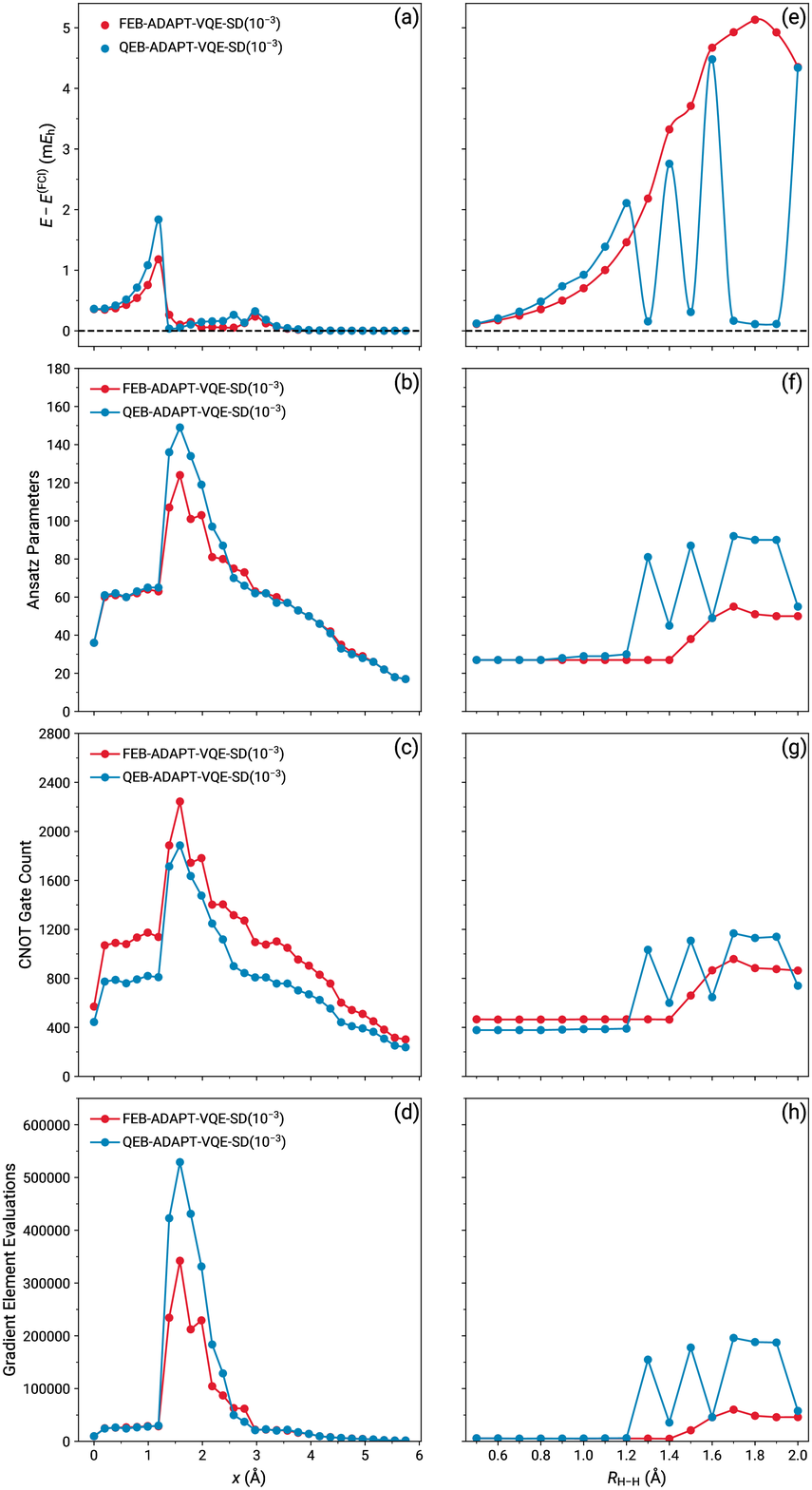}
	\caption{
		Errors relative to FCI [(a) and (e)], ansatz parameters [(b) and (f)], CNOT gate counts 
		[(c) and (g)], and gradient element evaluations [(d) and (h)] characterizing the FEB- and 
		QEB-\adaptsd{3} simulations of the insertion of Be to $\text{H}_2$ (left 
		column) and the symmetric dissociation of the $\text{H}_6$ ring (right column) as 
		described by the STO-6G basis.
	}
\end{figure*}

\begin{figure*}[h!]
	\centering
	\includegraphics[height=0.85\textheight]{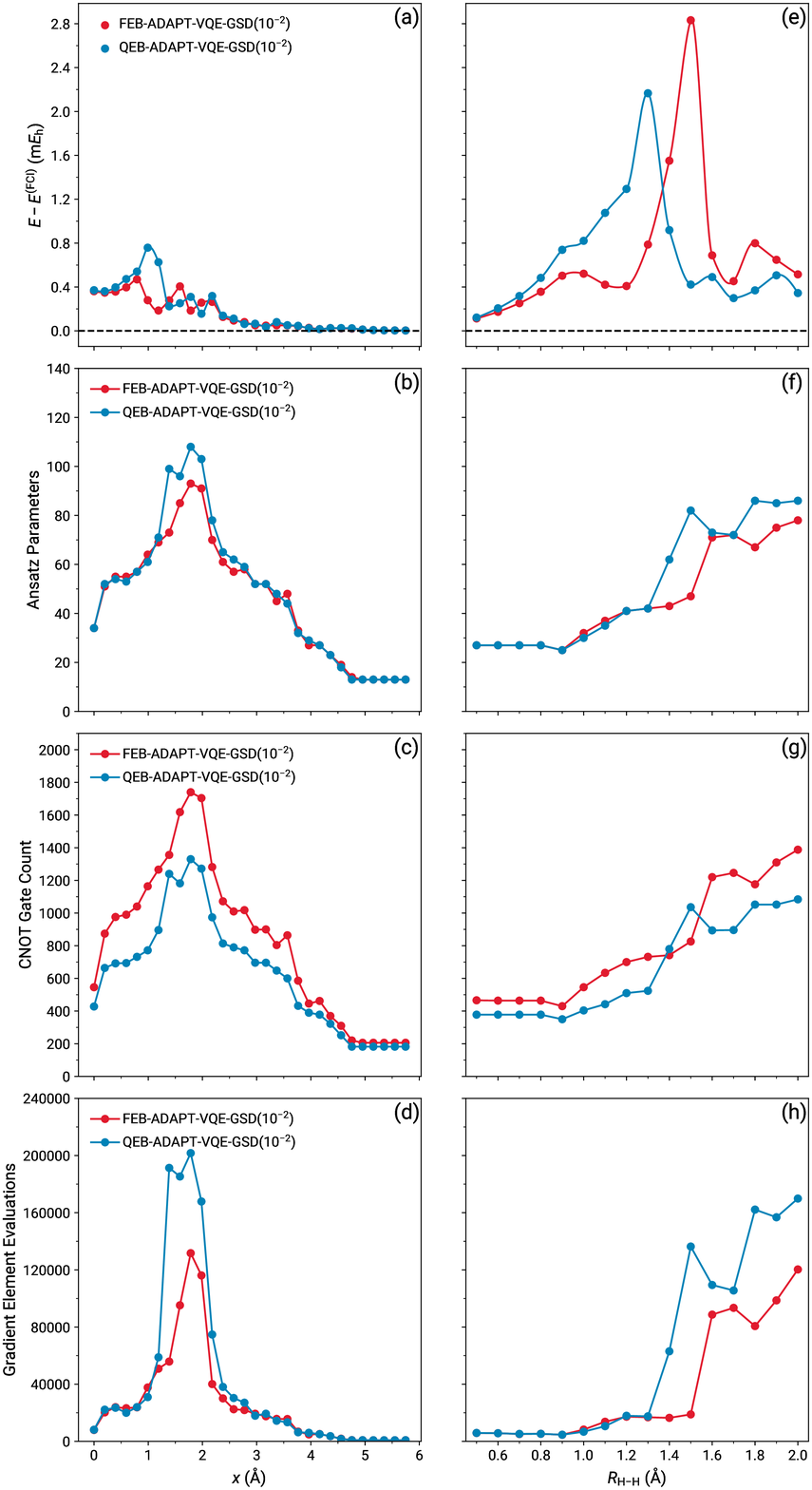}
	\caption{
		Errors relative to FCI [(a) and (e)], ansatz parameters [(b) and (f)], CNOT gate counts 
		[(c) and (g)], and gradient element evaluations [(d) and (h)] characterizing the FEB- and 
		QEB-\adaptgsd{2} simulations of the insertion of Be to $\text{H}_2$ (left 
		column) and the symmetric dissociation of the $\text{H}_6$ ring (right column) as 
		described by the STO-6G basis.
	}
\end{figure*}

\begin{figure*}[h!]
	\centering
	\includegraphics[height=0.85\textheight]{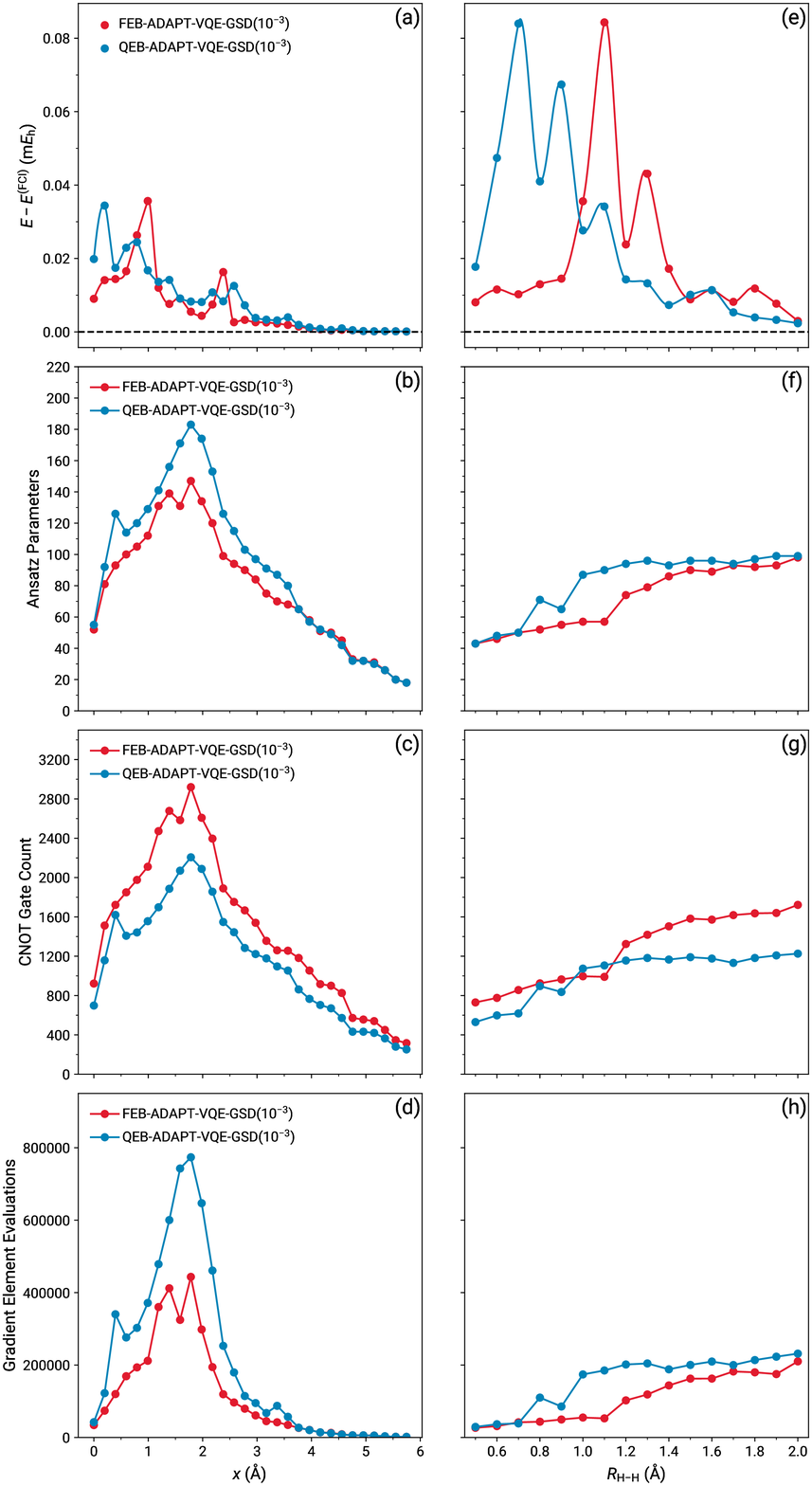}
	\caption{
		Errors relative to FCI [(a) and (e)], ansatz parameters [(b) and (f)], CNOT gate counts 
		[(c) and (g)], and gradient element evaluations [(d) and (h)] characterizing the FEB- and 
		QEB-\adaptgsd{3} simulations of the insertion of Be to $\text{H}_2$ (left 
		column) and the symmetric dissociation of the $\text{H}_6$ ring (right column) as 
		described by the STO-6G basis.
	}
\end{figure*}
\FloatBarrier

\subsection{FEB-SPQE \foreign{vs} QEB-ADAPT-VQE}
\label{sm_sec_feb-spqe_vs_qeb-adapt}

\begin{figure*}[h!]
	\centering
	\includegraphics[height=0.75\textheight]{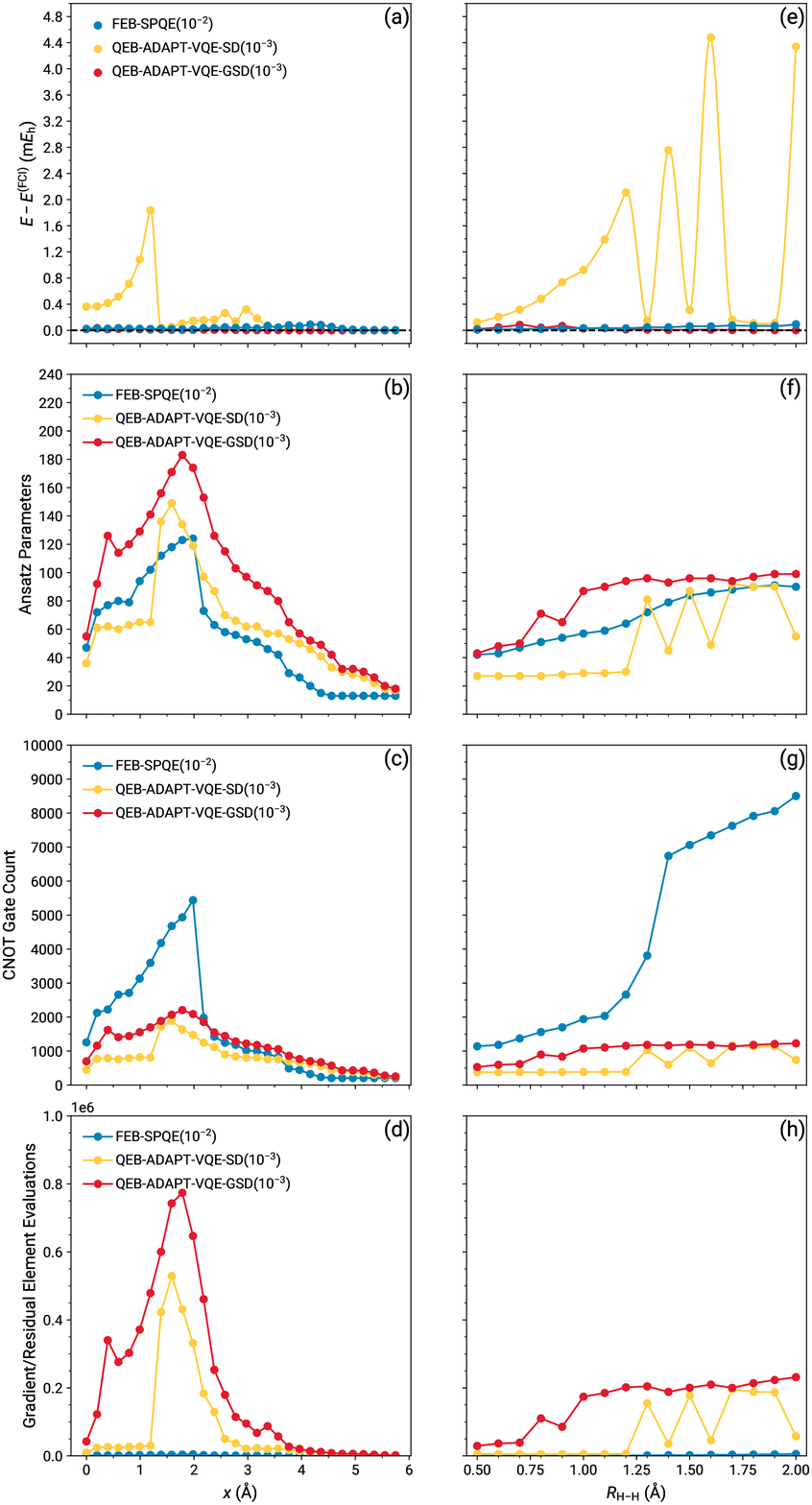}
	\caption{
		Errors relative to FCI [(a) and (e)], ansatz parameters [(b) and (f)], CNOT gate counts 
		[(c) and (g)], and gradient/residual element evaluations [(d) and (h)] characterizing 
		the FEB-\spqe{2}, QEB-\adaptsd{3}, and QEB-\adaptgsd{3} simulations of the 
		insertion of Be to $\text{H}_2$ (left column) and the symmetric 
		dissociation of the $\text{H}_6$ ring (right column) as described by the STO-6G basis.
	}
\end{figure*}

\end{document}